\documentclass[aps,prresearch,twocolumn,groupedaddress,longbibliography]{revtex4-2}

\usepackage{graphicx}
\usepackage{comment}
\usepackage{amsmath,amssymb}
\usepackage{xcolor}
\usepackage{dsfont}
\usepackage{hyperref}
\usepackage{blindtext}
\usepackage{microtype}
\usepackage{physics}

\usepackage[a4paper,margin=1in]{geometry}

\hypersetup{
    colorlinks=true,
    linkcolor=blue,
    citecolor=blue,
    urlcolor=blue
}

\newcommand{\rev}[1]{{\color{black}#1}}
\begin{document}

\author{Marc Nairn}
\author{Beatriz Olmos}
\author{Parvinder Solanki}

\affiliation{\vspace{0.5em} Institut f\"ur Theoretische Physik, Universit\"at T\"ubingen, Auf der Morgenstelle 14, 72076 T\"ubingen, Germany}

\title{Controlling emergent dynamical behavior via phase-engineered strong symmetries}

\begin{abstract}
Symmetry constraints provide a powerful means to control the dynamics of open quantum systems. However, the set of accessible control parameters is often limited. Here, we show that a tunable phase in the collective light-matter coupling of a cavity QED system induces a phase-dependent strong symmetry of the Liouvillian, enabling dynamical control of the open quantum system evolution. We demonstrate that tuning this phase substantially reduces the critical driving strength for dissipative phase transitions between stationary and non-stationary phases. We illustrate this mechanism in two experimentally relevant cavity QED settings: a two-species ensemble of two-level atoms and a single-species ensemble of three-level atoms. Our results establish phase control as a versatile tool for engineering dissipative phase transitions, with implications for quantum state preparation.
\end{abstract}

\date{\today}

\maketitle


\section{Introduction}

\begin{figure*}
    \centering
\includegraphics[width=\textwidth]{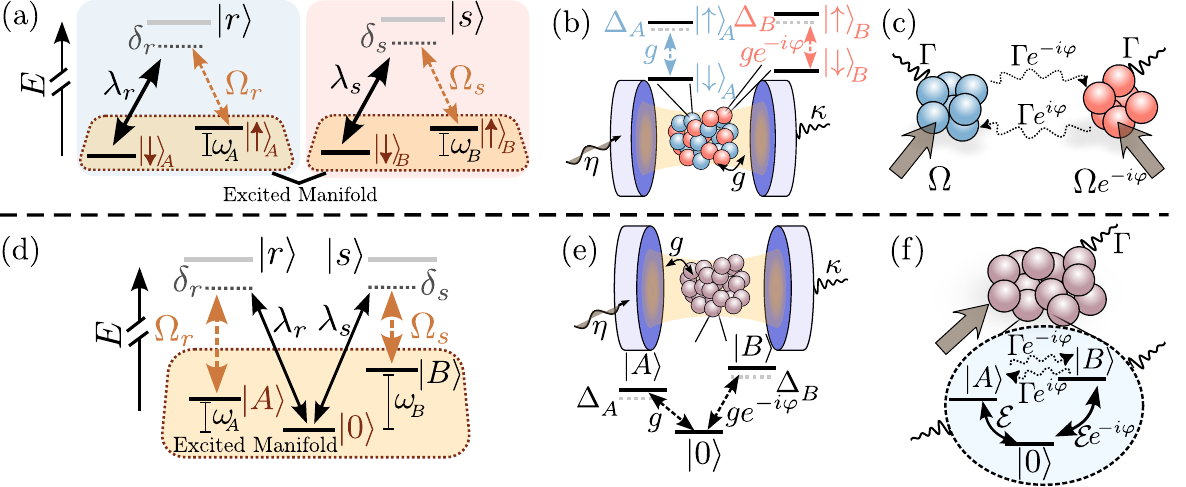}
\caption{\textbf{Schematic of setups.} \rev{\textbf{(a)} Two species ($A, B$) of multilevel atoms with distinct ground states; transitions to far-detuned states $\ket{r,s}$ are driven by $\Omega_{r,s}$, while the cavity mediates photon exchange between excited states via $\lambda_{r,s}$. \textbf{(b)} Setting $\Omega_{r,s} = |\Omega|e^{i\varphi_{r,s}}$ with $\varphi_r=0$ and $\varphi_s\equiv\varphi$, the low-energy dynamics reduce to a Tavis-Cummings model with species-dependent couplings $g_A=g$ and $g_B=ge^{-i\varphi}$. \textbf{(c)} Tracing away the cavity field yields an effective model of spin-only interactions mediated by the common photonic mode. \textbf{(d--f)} Equivalent configurations for a shared ground state system, where each atom is modeled as a three-level system: \textbf{(d)} the multilevel excitation scheme, \textbf{(e)} the resulting state-dependent couplings, and \textbf{(f)} the effective atom-only description.}}
    \label{fig:fig1}
\end{figure*}
Symmetries play a central role in modern physics by constraining dynamics, enforcing conservation laws, and shaping phases of matter and their transitions \cite{Kosmann:2011, Strocchi:2005}.
In open quantum systems \cite{Breuer:2002, Alicki:2007}, where coherent and dissipative processes coexist, symmetries impose stringent constraints on both the structure of the dynamics and the stationary and non-stationary states that they support \cite{Buca:2012,Albert:2014}. In these systems, when present, so-called strong symmetries remain preserved under both coherent evolution and dissipation, enforcing microscopic conservation laws in contrast to the ensemble-averaged conservation laws associated with weak symmetries \cite{Baumgartner:2008}. Such symmetries can give rise to decoherence-free subspaces \cite{Wu:2002, Beige:2000, Albert:2016, Lidar:2003, Shabani:2005, Bacon:2001, Vaecairn:2025} and enable rich non-equilibrium phenomena, including non-stationary phases
\cite{Iemini:2018, Alaeian:2022, Buca:2019, Kongkhambut:2022} with applications in sensing \cite{Cabot:2024a, Gribben:2025}, heat engines \cite{Carollo:2020, Yaletel:2023}, and energy storage \cite{Paulino:2025}.

A complementary route to controlling open-system dynamics exploits phase-engineered interference, achievable through geometric design \cite{Pleinert:2017} or reservoir engineering \cite{Metelmann:2015}.
Such phases can be global, generating collective nonlinear effects protected by symmetries \cite{Young:2025}, or appear as non-gaugeable relative phases that enable non-reciprocal behavior \cite{Fruchart:2021,Clerk:2022,Barzanjeh:2025}. These mechanisms underpin applications ranging from directional amplification \cite{Wanjura:2023, Jing:2021, Vega:2025} to the emergence of novel dynamical phases \cite{Chiacchio:2023, Nadolny:2025, Lyu:2025,ho2025highersymmetrybreakingnonreciprocity}.
Despite these advances, the use of phase-dependent symmetries themselves as a resource for controlling open-system dynamics has remained largely unexplored.

In this work, we show that the interplay between strong symmetries and tunable phases provides a powerful and systematic route to controlling dynamical phenomena in collective light-matter systems. We consider a driven-dissipative cavity QED platform that exhibits dissipative phase transitions between stationary and non-stationary states, also known as continuous-time crystals \cite{Iemini:2018,Buca:2019,PhysRevLett.123.260401,pizzi2019periodn,hurtado2020raretimecrystal,Carollo:2022,seeding2022michal,krishna2022measurement,PhysRevA.107.L010201,Iemini:2024,Solanki:2024,cabot2023nonequilibrium,mattes2023entangled,Mukherjee2024correlations,igor_thermodynamics,Paulino:2025,PhysRevA.108.023516,dc2s-94gv,jhd4-1khw,vnt1-m777,schumann2026hierarchicaltimecrystals}.
We demonstrate that a tunable phase in the collective light-matter coupling induces a phase-dependent strong symmetry that reshapes the onset and character of the non-stationary dynamics, enabling access to time-crystalline regimes at substantially reduced driving strengths.
We establish the generality of this mechanism by analyzing two variants of the Tavis-Cummings model: a two-species ensemble of two-level atoms and a single-species three-level atomic gas in a single-mode cavity.
Our results identify phase-controlled strong symmetries as a versatile tool for engineering non-stationary phases and controlling dissipative phase transitions in open quantum systems, with potential applications in quantum state preparation.

\section{Strong symmetries in open quantum systems} Within the Born-Markov approximation, the time evolution of the density matrix $\hat{\rho}$ of an open quantum system is governed by the Gorini-Kossakowski-Sudarshan-Lindblad master equation,
\begin{equation}
\dot{\hat{\rho}} = \mathcal{L}(\hat{\rho}) \equiv -\dfrac{\mathrm{i}}{\hbar}[\hat H, \hat{\rho}] + \sum_k \mathcal{D}[\hat{L}_k]\hat{\rho}, \label{eq:GKLS}
\end{equation}
where $\hat H$ is the Hamiltonian, $\hat L_k$ are the Lindblad jump operators, $\mathcal{D}[\hat L_k]\hat{\rho}=\hat{L}_k \hat{\rho} \hat{L}_k^\dagger - \{\hat{L}_k^\dagger \hat{L}_k, \hat{\rho}\}/2$ denotes the dissipator and $\mathcal{L}$ is the Liouvillian superoperator.

A system exhibits a \textit{strong symmetry} if there exists an operator $\hat A$ that commutes with both the Hamiltonian and all Lindblad jump operators, i.e., $[\hat A,\hat H]=0$ and $[\hat A,\hat L_k]=0\ \forall k$ \cite{Buca:2012, Baumgartner:2008, Albert:2014}.
This dual commutation relation partitions the Hilbert space into invariant symmetry sectors, $\mathcal H=\bigoplus_\lambda \mathcal H_\lambda$, corresponding to the eigenspaces of $\hat A$. As a result, the dynamics generated by $\mathcal{L}$ is decoupled across sectors. Decomposing the initial state as $\hat\rho(0)=\sum_{\lambda,\lambda'}\hat\rho_{\lambda\lambda'}(0)$ with $\hat\rho_{\lambda\lambda'}(0)=\hat{\Pi}_\lambda\hat\rho(0)\hat{\Pi}_{\lambda'}$ and $\hat\Pi_\lambda$ being the projector onto $\mathcal H_\lambda$, the coherences between different symmetry sectors ($\lambda\neq\lambda'$) do not contribute to the long-time dynamics \cite{Iemini:2024}. Consequently, the steady state of Eq.~(\ref{eq:GKLS}) assumes the block-diagonal form
\begin{equation}
\hat{\rho}^{\mathrm{ss}}
=\lim_{t\to\infty}e^{\mathcal L t}\hat\rho(0)
=\sum_\lambda w_\lambda\hat\rho^{\mathrm{ss}}_\lambda,
\label{eq:general_rhoss}
\end{equation}
where the conserved sector weights (or overlaps) $w_\lambda=\Tr[\hat\Pi_\lambda\hat\rho(0)]$ are fixed by the initial state.

In the following, we show that in collective cavity QED systems, a tunable phase in the light-matter coupling modifies the underlying strong symmetry, thereby reshaping the dynamical constraints imposed by Eq.~(\ref{eq:GKLS}) and controlling the onset of dissipative phase transitions.

\section{Two-species model} 

We first consider an ensemble of $N$ atoms interacting with a single-mode driven cavity. The constituents belong to two different species, $A$ and $B$, each comprising $N/2$ \rev{three-level atoms with states $\ket{\downarrow}_{A/B}$ and $\ket{\uparrow}_{A/B}$ coupled via a Raman transition to a far-detuned excited state, Fig.~\ref{fig:fig1}(a). \rev{As detailed in Appendix~\ref{app:microscopic}}, after adiabatic elimination of the off-resonant Raman channel, each species is described as a two-level system collectively coupled to the cavity field with coupling strengths \begin{equation}
    g_A = -\dfrac{\lambda_r\Omega^*_r}{2\delta_r}, \quad  g_B = -\dfrac{\lambda_s\Omega^*_s}{2\delta_s} 
\end{equation} and atomic detunings $\Delta_{A/B} = \delta_{A/B} - |\Omega_{r/s}|^2/4\delta_{r/s}$. Both the magnitude and phase of the couplings, as well as the detunings, can be controlled experimentally using external Raman lasers. In the following, we focus on the situation in which the two species differ only by a controllable phase $\varphi$ and let $\Delta=\Delta_A=\Delta_B$. To achieve this, we set the cavity transitions to be equal $\lambda_s=\lambda_r$ and imprint a relative phase onto the Raman lasers $\Omega_{r,s} = |\Omega|e^{i\varphi_{r,s}}$ with $\varphi_r=0$ and $\varphi_s\equiv\varphi$. The low-energy dynamics thus reduce to a Tavis-Cummings model with species-dependent couplings $g_A=g$ and $g_B=ge^{-i\varphi}$; see Fig.~\ref{fig:fig1}(b).

The resulting many-body Hamiltonian in the appropriate rotating frame reads
\begin{equation}
\begin{split}
    \hat{H}_{2} = \hbar\Delta_c\hat{a}^\dagger\hat{a} 
    + i\hbar\sqrt{N}\eta\left(\hat{a}^\dagger - \hat{a}\right)  
    &+ \hbar\Delta\hat{S}_{\varphi}^z\\
        +\,\frac{\hbar g}{\sqrt{N}}\left(\hat{a}^\dagger\hat{S}_\varphi 
    + \hat{a}\hat{S}^\dagger_\varphi\right)&,
    \label{eq:H2}
    \end{split}
\end{equation}
}
where $\hat{a}$ ($\hat{a}^\dagger$) annihilates (creates) a cavity photon, $\eta$ is the drive amplitude, $\Delta_c$ the cavity detuning, and $\hat{S}_\varphi = \sum_{j=1}^{N/2}\left(\hat{\sigma}_{jA} + 
e^{i\varphi}\hat{\sigma}_{jB}\right)$ with $\hat{\sigma}_{jm} = \ket{\downarrow}\bra{\uparrow}_\rev{{jm}}$ is the phase-dependent collective lowering operator. The associated $z$-component is $\hat{S}_\varphi^z = 
\frac{1}{2}\left(\hat{S}_\varphi^\dagger\hat{S}_\varphi - \hat{S}_\varphi\hat{S}_\varphi^\dagger\right)$. Photon losses are described by a Lindblad dissipator with decay rate $\kappa$, and all terms are scaled with $N$ to ensure a well-defined thermodynamic limit ($N\to\infty$). Throughout this work we set $\Delta_c=0$; qualitatively similar behavior persists for $\Delta_c\neq 0$ (\rev{see Appendix~\ref{app:cavity_detuning}}).

To understand the role of the phase $\varphi$, we first analyze the mean-field dynamics in terms of the cavity field $\alpha = \langle\hat{a}\rangle$ and the rescaled collective spin components $s_{A/B}^{\mu}=\langle \hat{S}_{A/B}^{\mu} \rangle/(N/2)$ with $\hat{S}_{m}^\mu = \frac{1}{2}\sum_{j=1}^{N/2}\hat{\sigma}_{jm}^\mu$ and $\hat{\sigma}_{jm}^{\mu=x,y,z}$ the Pauli matrices for $m\in\{A,B\}$. To distinguish stationary from non-stationary phases, we introduce the dominant frequency $\tilde{\omega}$ extracted from the spectrum of the spin polarization \rev{$\mathcal{F}_{A/B}(\tilde\omega) = \int_{-\infty}^\infty s^z_{A/B}(t) e^{-i\tilde\omega t}dt$}. In all cases considered, the two species oscillate synchronously, so that $\tilde\omega\equiv\tilde\omega_A = \tilde\omega_B$.

For $\varphi=0$ and $\Delta=0$, the system undergoes a transition from a stationary state ($\tilde{\omega}=0$) to a non-stationary state ($\tilde{\omega}>0$) at a critical driving strength $\eta_c=g$, as shown in Fig.~\ref{fig:fig2}(a). In contrast, for finite $\varphi$ the critical driving strength becomes phase-dependent for any initial state with $\vert s_{A/B}^{z}\vert < 1$.
This behavior is illustrated in Fig.~\ref{fig:fig2}(a) for the initial state condition $(\alpha, s^x_{A/B}, s^y_{A/B}, s^z_{A/B})=(0,1,0,0)$.
Remarkably, the threshold for the onset of non-stationarity is progressively reduced with increasing $|\varphi|$ and vanishes in the limit $\varphi \to \pm \pi$.

In the following section, we show that this phase-enabled reduction of the critical drive originates from a modification of an underlying strong symmetry, which reshapes the accessible dynamical sectors of the system.

\subsection{Symmetry characterization}
In the limit of strong dissipation $\kappa \gg g,\eta,\Delta$, the cavity field can be adiabatically eliminated, yielding an effective spin-only description (\rev{see Appendix~\ref{app:adiabatic}}), Fig.~\ref{fig:fig1}(c).
The reduced spin density matrix $\hat{\mu}_{s}$ then evolves according to
\begin{equation}
\dot{\hat{\mu}}_s = -\frac{\mathrm{i}}{\hbar}[\hat{H}_S,\hat{\mu}_{s}] +\frac{\Gamma}{2S}\mathcal{D}[\hat{S}_\varphi]\hat{\mu}_s, \label{eq:2LS_spin_only}
\end{equation}
where $S=N/2$, $\Gamma = 4g^2/\kappa$, and the effective Hamiltonian reads
$
\hat H_S =\hbar\Delta\hat{S}_{\varphi}^z +\hbar \Omega\left(\hat S_\varphi + \hat S^\dagger_\varphi\right)
$
with $\Omega = 2\eta g/\kappa$.
\rev{Here, collective dissipation $\mathcal{D}[\hat{S}_\varphi]\hat{\mu}_s$ induces all-to-all interactions between the spin ensembles.}
For $\Delta=0$ and $\varphi=0$, Eq.~\eqref{eq:2LS_spin_only} reduces to the boundary time crystal (BTC) model \cite{Iemini:2018}, which undergoes a transition from a stationary to a time crystalline phase at $\Omega/(\Gamma/2)=1$.
This directly accounts for the critical driving strength $\eta_c=g$ observed in the full cavity-spin model and motivates our identification of the non-stationary phase as a continuous time crystal.

To elucidate the role of the phase $\varphi$, we analyze the symmetries of the spin-only dynamics. The operators $\hat{S}_\varphi, \hat{S}_\varphi^\dagger, \hat{S}_\varphi^z$ generate an $\mathfrak{su}(2)$ algebra. The associated Casimir operator $\hat{\mathbf{S}}_\varphi^2 = \dfrac{1}{2}\left(\hat{S}_\varphi\hat{S}_\varphi^\dagger+\hat{S}_\varphi^\dagger\hat{S}_\varphi\right)+(\hat{S}_\varphi^z)^2$ commutes with both the Hamiltonian $\hat H_S$ and the jump operators $\hat S_\varphi$, and therefore constitutes a strong symmetry of the Liouvillian. This symmetry generalizes a total spin conservation in the $\varphi=0$ case to a phase-dependent setting. As a consequence, the dynamics decomposes into invariant sectors labeled by the eigenvalues $S^2_\varphi$ of $\hat{\mathbf{S}}_\varphi^2$. The system can be thus described in a generalized, phase-dependent Dicke basis. For $\varphi=0$, these sectors coincide with the standard Dicke manifolds  \cite{Shammah:2018, Dicke:1954}; for $\varphi\neq0$, they correspond to rotated collective modes encoding the relative phase between the two subensembles. The weight of each sector is $w_{S^2}^\varphi = \mathrm{Tr}[\hat{\Pi}^\varphi_{S^2}\hat{\mu}_s]$, where $\hat{\Pi}^\varphi_{S^2}$ projects onto the corresponding eigenspace. In practice, the dynamics is governed by the competition between the fully symmetric sector, $w_{{S^2}_{\mathrm{max}}}^\varphi$, and its complement, $\overline{w}^\varphi = 1 - w_{{S^2}_{\mathrm{max}}}^\varphi$, which collects all contributions outside the permutationally invariant Dicke manifold.

\subsection{Dynamical phases in the two species model} \label{sec:Dyn Phase spin 1/2} 
\begin{figure}
    \centering
    \includegraphics[width=0.49\textwidth]{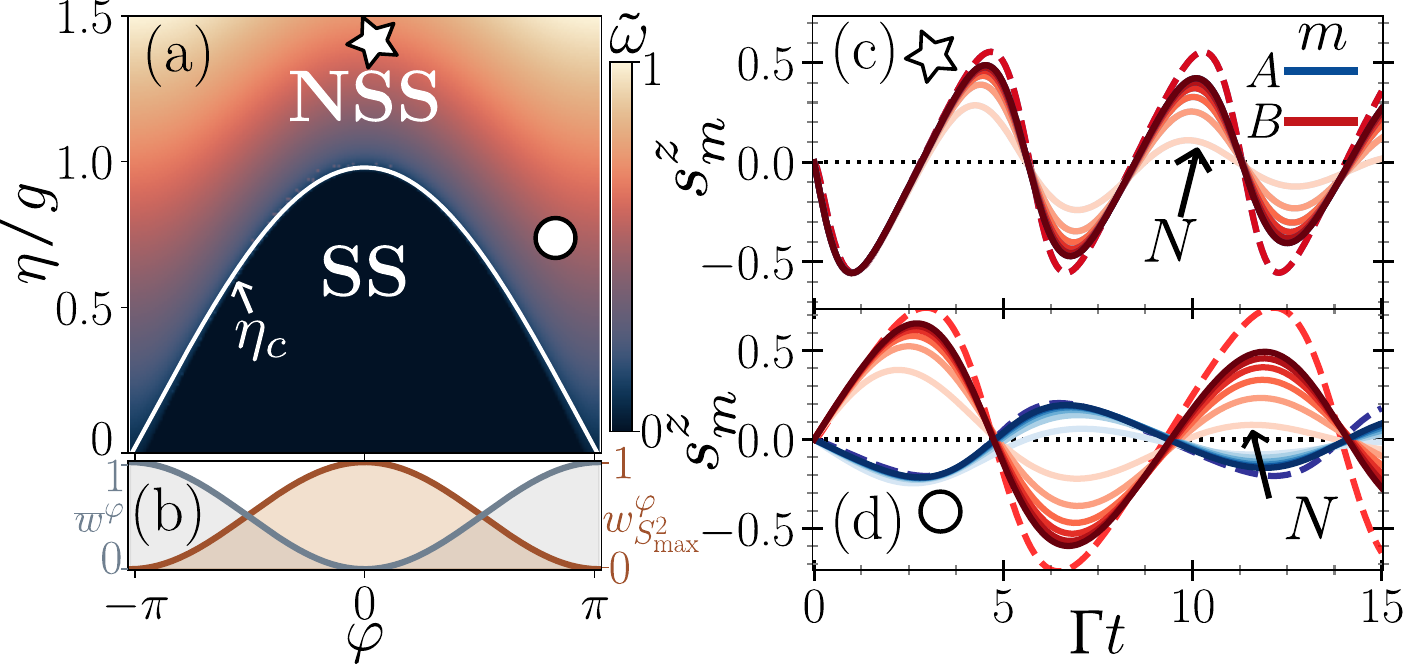}
    \caption{\textbf{Phase enabled nonstationary states}. \textbf{(a)} Mean-field phase diagram showcasing the effect of tunable phase $\varphi$ on the transition between stationary (\textbf{SS}) and nonstationary states (\textbf{NSS}), and \textbf{(b)} the corresponding weight distribution in the Dicke subspace, $w_{S^2_\text{max}}^\varphi$ and its complement $\overline{w}^\varphi$.   Finite-size dynamics for highlighted points in \textbf{(a)} illustrating the usual time-crystalline behavior and the lowering of the threshold value, panels \textbf{(c)} and \textbf{(d)}, respectively. We consider system sizes $N\in\left(10,20,\dots,60\right)$ (lines with increasing color intensity). The results are benchmarked against the mean-field values (dashed) with   $g/\kappa=0.1$ and $\Delta=0$.}
    \label{fig:fig2}
\end{figure}

We now connect this symmetry structure to the dynamical phase diagram shown in Fig.~\ref{fig:fig2}(a). We consider the initial product state $\hat{\mu}_{s}(0) = \bigotimes_{m=A,B} \bigotimes_{j=1}^{N/2} \ketbra{+_x}_{jm}$ with $\ket{+_x} = \left(\ket{\uparrow} + \ket{\downarrow}\right)/\sqrt{2}$, which corresponds to the mean-field condition used in Fig.~\ref{fig:fig2}(a). While this choice is not unique, it is particularly instructive because its decomposition into symmetry sectors depends sensitively on $\varphi$.

\begin{figure*}[t!]
    \centering
    \includegraphics[width=0.85\linewidth]{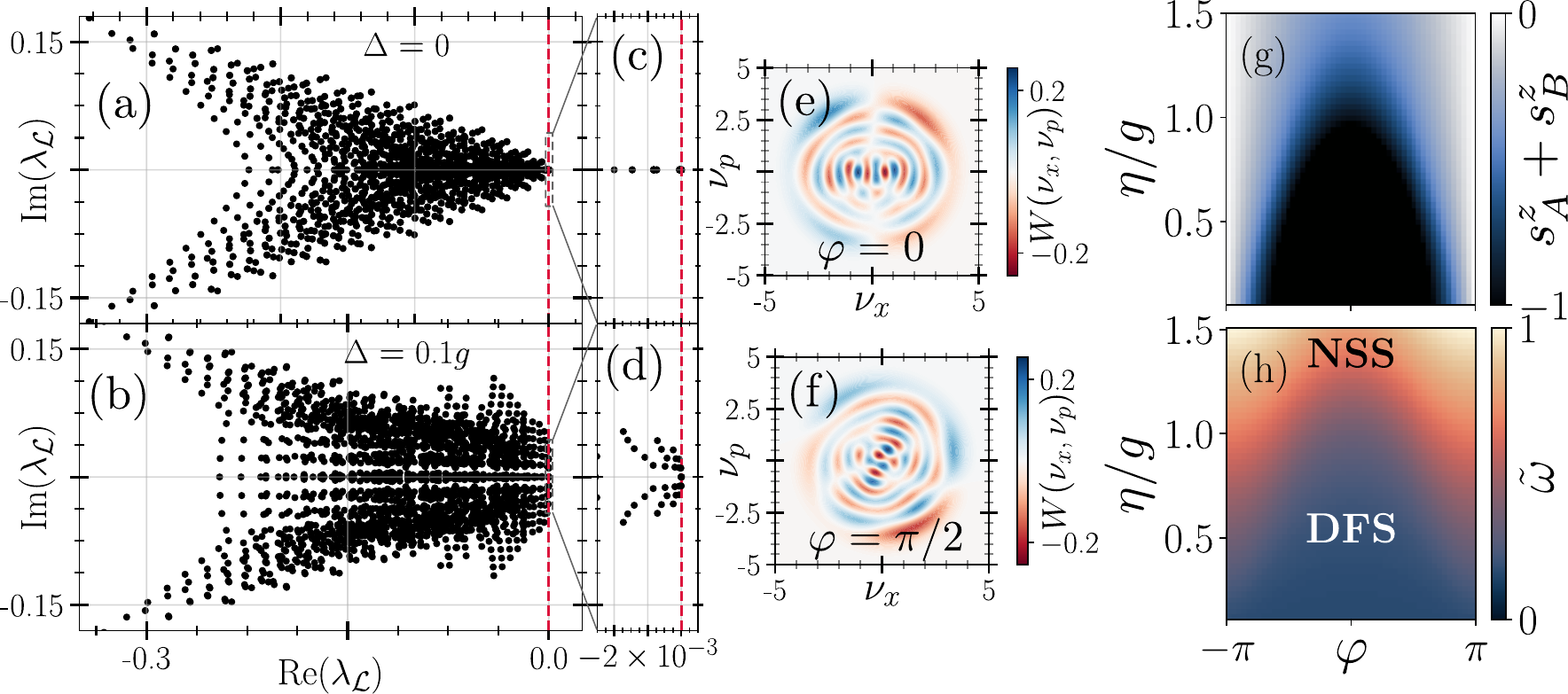}
    \caption{\textbf{Role of spin detuning and relative phase on spin-1/2 Liouvillian spectrum}. Liouvillian eigenvalue ($\lambda_\mathcal{L}$) spectrum for $N_A=N_B=3$ for equal spin detunings $\Delta_A=\Delta_B\equiv\Delta$, for $\Delta=0$ \textbf{(a)} and $\Delta=0.1g$ \textbf{(b)}. Zoomed-in regions close to the $\mathrm{Re}(\lambda_\mathcal{L})=0$ line highlight the nature of the long-time behavior. Besides the unique $\mathrm{Re}(\lambda_\mathcal{L})= \mathrm{Im}(\lambda_\mathcal{L})=0$ steady state, we find purely decaying modes ($\mathrm{Re}(\lambda_\mathcal{L})<0$) in \textbf{(c)} and the existence of purely imaginary eigenvalues $\mathrm{Im}(\lambda_\mathcal{L})= \pm\Delta$ in \textbf{(d)}, indicators of a decoherence-free subspace. The phase $\varphi$ does not contribute to the eigenvalue spectrum, but it can be used to rotate the eigenvectors of the Liouvillian. In \textbf{(e)}-\textbf{(f)} we plot the Wigner function of the eigenvector $\ket{\nu_+}$ for $N_A=N_B=1$, corresponding to the $\lambda_+=i\Delta$ mode, for two choices of $\varphi$. In \textbf{(g)--(h)} we map out the resulting mean-field phase diagram for nonzero detunings $\Delta=0.1g$, showing both the total magnetization and the dominant normalized frequency $\tilde{\omega}_A$ used to discern between a Decoherence-Free Subspace (\textbf{DFS}) and Nonstationary States (\textbf{NSS}). Other parameters: $\eta/\kappa=0.025$, $\Delta_c=0$, $g/\kappa=0.1$.}
    \label{fig:fig_dfs_2s}
\end{figure*}

For $\varphi=0$, the initial state lies entirely in the fully symmetric Dicke manifold, $w_{{S^2}_{\mathrm{max}}}^{\varphi=0}=1$, and the critical driving strength reaches its maximal value, reproducing the BTC threshold \cite{Iemini:2018, Carollo:2022}. As $\varphi$ is tuned away from zero, weight is transferred into non-symmetric sectors, as shown in Fig.~\ref{fig:fig2}(b). Since different symmetry sectors are dynamically decoupled, only the component within the fully symmetric manifold undergoes the collective instability responsible for time-crystalline dynamics. The effective evolution is therefore governed by a Dicke state with reduced effective spin length, leading to a corresponding reduction of the critical driving strength \cite{Solanki:2024, Cabot:2024b}. \rev{We stress that this effect is markedly different from what would be observed in the case of coexisting fixed-point and limit-cycle attractors \cite{Bhaseen:2012, Hotter:2023, Mivehvar:2024} as here the initial-state dependence arises from the permanent decoupling of symmetry sectors, each hosting a unique attractor, rather than from competition between distinct basins of attraction}. 

In the limit of $\varphi\to\pm\pi$, the contribution of the fully symmetric Dicke manifold vanishes $w_{{S^2}_{\mathrm{max}}}^\varphi\to0$. In this regime, arbitrarily weak driving induces time-crystalline dynamics \cite{Iemini:2024, Solanki:2024}, consistent with the vanishing threshold observed in Fig.~\ref{fig:fig2}(a). At the exact antisymmetric point (exactly $\varphi=\pm\pi$), the initial state becomes a dark state, remaining unaffected by the dissipative dynamics. Within the time-crystalline phase, both species exhibit frequency-locked oscillations that converge to the mean-field behavior as $N$ increases, Fig.~\ref{fig:fig2}(c,d). The oscillation frequency $\tilde\omega$ increases with the weight $\overline{w}^\varphi$, reflecting the contribution of dressed modes originating outside the fully symmetric manifold \cite{Iemini:2024}.

\rev{

\subsection{Decoherence-free subspaces in the two-species model} We now turn to the role of finite spin detuning, $\Delta \neq 0$, which qualitatively alters the dynamics and induces decoherence-free subspaces (DFS).

\begin{figure*}[t]
    \centering
    \includegraphics[width=0.9\textwidth]{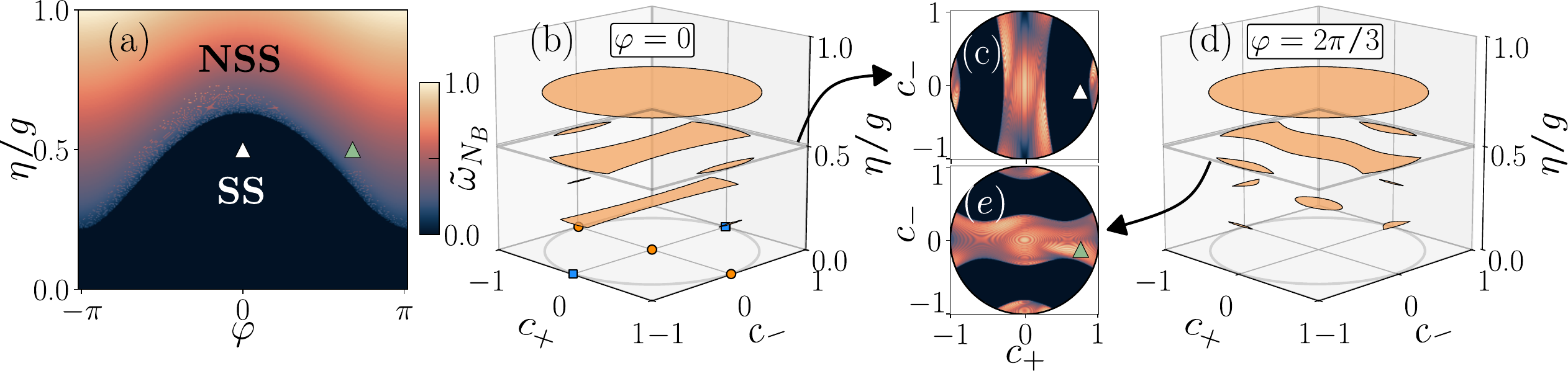}
    \caption{\textbf{Three-level system frequency response}. \textbf{(a)} Phase-dependent steady-state phase diagram showcasing frequency response for a chosen initial state with $c_\pm=(1\pm\sqrt2)/2\sqrt2, c_0=\sqrt{1-c_+^2-c_-^2}$, far from the eigenstates $\ket{v_{\pm1}^\varphi}$. The order parameter $\tilde\omega_{N_B}$ is the normalized dominant Fourier component of the excited state $B$. \textbf{(b)} Initial state visualization of threshold contours in $(c_+,c_-)$ parameter space for $\varphi=0$. States are defined as nonstationary if $\tilde\omega_{N_B}\geq0.01$ and shaded orange. \rev{This threshold is chosen to exceed the numerical noise floor, and the phase boundaries remain qualitatively unchanged for variations within an order of magnitude}. On the $\eta/g=0$ plane we highlight the eigenstates of $\hat{\tau}_j^\varphi$ mentioned in the main text.
    \textbf{(c)} Parameter space map taken at the $\eta/g=0.5$ slice  from \textbf{(b)}. \textbf{(d)} Threshold contours for $\varphi=2\pi/3$, and the equivalent parameter space map \textbf{(e)}. The white and green markers correspond to the initial state used in \textbf{(a)}, falling inside the nonstationary regime as $\varphi$ is tuned. Other parameters used $g/\kappa=0.1, \Delta=0$, for a simulation time of $g t=2\times10^4$.}
    \label{fig:fig3}
\end{figure*}

Since $\mathcal{L}$ is a non-Hermitian superoperator, it exhibits distinct left and right eigenvectors; we denote these as $\nu_\alpha$ and $\mu_\alpha$, respectively, defined by
\begin{equation}
\mathcal{L}\nu_\alpha = \lambda_\alpha \nu_\alpha,
\qquad
\mathcal{L}^\dagger\mu_\alpha  = \lambda_\alpha^* \mu_\alpha,
\end{equation}
satisfying the biorthogonality condition $\mathrm{Tr}\left(\mu_\alpha^\dagger \nu_\beta\right) = \delta_{\alpha\beta}$. The dynamical map then admits the expansion \cite{Minganti:2018}
\begin{equation}
e^{\mathcal{L} t}[\circ]
= \sum_\alpha e^{\lambda_\alpha t}
\nu_\alpha \mathrm{Tr}\left(\mu_\alpha^\dagger \circ\right),
\end{equation}
where the real part of $\lambda_\alpha$ sets the decay rate and its imaginary part sets the oscillation frequency of mode $\nu_\alpha$.

For $\Delta=0$, the spectrum of $\mathcal{L}$ lies entirely on the negative real axis at finite $N$, so all oscillations are damped. Persistent oscillations (non-stationary states) emerge only in the thermodynamic limit $N\to\infty$ when the driving is sufficiently strong, as discussed above. When an equal and finite detuning $\Delta = \Delta_A = \Delta_B \neq 0$ is introduced, a pair of purely imaginary eigenvalues $\lambda_\pm = \pm i\Delta$ appears even at finite $N$, as shown in Fig.~\ref{fig:fig_dfs_2s}(a)--(d). The associated eigenmodes are unaffected by the dissipators and form a decoherence-free subspace. Consequently, for $\eta < \eta_c$ the system does not relax to a stationary state but instead exhibits persistent oscillations at frequency $\Delta$. For $\eta > \eta_c$, the oscillation frequencies are modified in a manner analogous to the resonant case.

The phase $\varphi$ does not affect the eigenvalue spectrum, but it rotates the eigenvectors of the Liouvillian. To see this, note that the contribution of the DFS pair to the expectation value of any observable $O$ is
\begin{equation}
\langle O\rangle_\pm(t) = e^{\pm i\Delta t} A_O(\varphi),
\label{eq:dfs_obs}
\end{equation}
where
\begin{equation}
\begin{split}
A_O(\varphi) &\equiv c_\pm(\varphi)\,\mathrm{Tr}\bigl(O\,\nu_\pm(\varphi)\bigr),\\
c_\pm(\varphi) &= \mathrm{Tr}\bigl(\mu_\pm(\varphi)^\dagger\rho(0)\bigr).
\end{split}
\end{equation}

Consequently, $\Delta$ fixes the oscillation frequency while the amplitude $A_O(\varphi)$ depends explicitly on $\varphi$ through the left and right eigenmodes. Changing $\varphi$ rotates the decoherence-free eigenvectors in Liouville space, modifying the overlaps and thereby enabling or suppressing oscillatory behavior without changing the frequency. This mode rotation is directly visible in the Wigner function plots of $\nu_+(\varphi)$, shown in Fig.~\ref{fig:fig_dfs_2s}(e)--(f) for $N_A=N_B=1$. The mean-field phase diagram for $\Delta \neq 0$ is shown in Fig.~\ref{fig:fig_dfs_2s}(g)--(h), where DFS regions are identified by extended $\tilde{\omega}_A = \mathrm{const.} \neq 0$ regions that persist across the driving axis, with their frequency fixed by $\Delta$. As the driving amplitude is increased, the system transitions into the nonstationary regimes discussed in Sec. \ref{sec:Dyn Phase spin 1/2}.
}

\section{Single-species three-level system} To 
illustrate the generality of phase-controlled strong symmetries, we now 
consider a single species of $N$ atoms, each with a common ground state 
$\ket{0}$ and two excited states $\ket{A}$ and $\ket{B}$, coupled to the 
same single-mode driven cavity, see Fig.~\ref{fig:fig1}(d). \rev{This 
configuration is the single-species analogue of the two-species setup: 
the same Raman scheme applies, with the two legs $A$ and $B$ now 
corresponding to distinct excited states of the same atom rather than 
two separate species. Imprinting a relative phase $\varphi$ onto the 
Raman drives as before yields $g_A = g$ and $g_B = ge^{-i\varphi}$ 
[see Fig.~\ref{fig:fig1}(e)], and the full many-body Hamiltonian takes 
the form
\begin{equation}
\begin{split}
    \hat{H}_{3T} = \hbar\Delta_c\hat{a}^\dagger\hat{a} 
    + i\hbar\sqrt{N}\eta\left(\hat{a}^\dagger - \hat{a}\right) 
    &+ \hbar\sum_{k}\Delta_k\hat{N}_k \\
    + \frac{\hbar g}{\sqrt{N}}\left(\hat{a}\hat{\Lambda}^\dagger_\varphi 
    + \hat{a}^\dagger\hat{\Lambda}_\varphi\right)&,
    \label{eq:H3T}
\end{split}
\end{equation}
where $\hat{\Lambda}_\varphi = \hat{\Lambda}_A + e^{i\varphi}\hat{\Lambda}_B$ with $\hat{\Lambda}_k = \sum_{j=1}^N\ketbra{0}{k}_j$ is the phase-dependent collective lowering operator, $\hat{N}_k = \hat{\Lambda}_k^\dagger\hat{\Lambda}_k$ counts the population in excited leg $k$, and $\Delta_k$ are the atomic 
detunings (see Appendix~\ref{app:microscopic} for the full derivation).} After adiabatic elimination of the photonic field in the bad-cavity limit $\kappa \gg g, \eta$ [see Fig.~\ref{fig:fig1}(f)], the atomic dynamics 
is governed by\begin{equation}
    \dot{\hat{\mu}}_{\rm 3s} = -\frac{i}{\hbar}[\hat{H}_{\rm 3s}, 
    \hat{\mu}_{\rm 3s}]+\frac{\Gamma}{N} 
    \mathcal{D}[\hat{\Lambda}_\varphi]\hat{\mu}_{\rm 3s},
\end{equation}
where $\Gamma = 4g^2/\kappa$, and the effective Hamiltonian reads
\begin{equation}
\hat{H}_\mathrm{3s} = \hbar\sum_k\Delta_k\hat{\Lambda}_k^\dagger\hat{\Lambda}_k 
+ \hbar\mathcal{E}\left(\hat{\Lambda}_\varphi + \hat{\Lambda}_\varphi^\dagger\right) \label{eq:3LS_spinonly}\end{equation}
with $\mathcal{E} = 2\eta g/\kappa$.

\subsection{Symmetries in three-level system}
For equal detunings, $\Delta_A=\Delta_B=\Delta$, the dynamics admits a phase-dependent strong symmetry that can be written as a sum of single-atom operators, $\hat{\mathcal{T}}^\varphi = \sum_{j=1}^N\hat{\tau}^\varphi_j$, with $\hat{\tau}^\varphi_j = \ketbra{0}{0}_j + e^{i\varphi}\ketbra{A}{B}_j + e^{-i\varphi}\ketbra{B}{A}_j$. To elucidate the structure of the dynamically invariant subspaces, we examine the eigenstates of the single-atom operators $\hat{\tau}^\varphi_j$, which split according to their eigenvalues $\pm1$:
\begin{equation}
      \ket{v_{+1}^\varphi}_j \equiv \big\{\ket{0}_j, \ket{+}_j\big\}, \ket{v_{-1}^\varphi}_j \equiv\ket{-}_j,
\end{equation}
where $\ket{\pm}_j = \left(\ket{B}_j \pm e^{i\varphi}\ket{A}_j\right)/\sqrt{2}$. For $\Delta=0$, the states in $\ket{v_{+1}^\varphi}_j$ form a doubly-degenerate subspace in which excitations can cycle between $\ket{0}_j$ and $\ket{+}_j$ under the combined action of Hamiltonian and dissipative dynamics. We therefore identify this as the \emph{bright subspace}. In contrast, $\ket{v_{-1}^\varphi}_j$ is a \emph{dark state}, decoupled from both coherent and dissipative dynamics. Extending this structure to $N$ atoms suggests that the population of dark states strongly influences the long-time dynamics. In particular, the fraction of atoms occupying the dark subspace, together with the driving rate $\eta$, determines whether the system evolves toward a stationary or non-stationary phase.

\subsection{State-space dependent dynamics} To illustrate the dynamical role of the bright and dark subspaces, we consider a family of pure initial states $\hat\mu_{3s}(0) = \ketbra{\psi_{3s}}$ with $\ket{\psi_{3s}} = c_+\ket{+}^{\otimes N} + c_-\ket{-}^{\otimes N} + c_0\ket{0}^{\otimes N}$, where $c_0 = \sqrt{1-c_+^2-c_-^2}$ and $c_\pm\in\mathbb{R}$ with $c_+^2+c_-^2\leq 1$. We quantify the support of a state on the symmetry sectors via the weights $w_\pm^\varphi=\mathrm{Tr}[\hat{\Pi}_\pm^\varphi\hat{\mu}_{3s}]$, where the projectors onto the bright and dark subspaces are $\hat{\Pi}_\pm^\varphi= \bigotimes_{j}^N \ket{v_{\pm1}^\varphi} \bra{v_{\pm1}^\varphi}_j$.

Non-stationary dynamics can arise either from sufficiently strong driving or from finite overlap with the freely evolving bright subspace ($w_+^\varphi>0$). 
\rev{The competition between drive and sector weight determines the long-time behavior, such that the critical driving strength $\eta_c$ depends on the effective spin length which can be actively controlled by tuning the phase $\varphi$. }
To characterize the non-stationary phase, we analyze the mean-field dynamics with initial conditions determined by expectation values in $\ket{\psi_{3s}}$. As an order parameter we use the normalized dominant frequency $\tilde\omega_{N_B}$ of the Fourier transform of the $B$-excited-state population. The choice of observable is not essential, and other choices yield equivalent results \cite{Prazeres:2021}.

For initial states with $0<w_\pm^\varphi<1$, the behavior closely parallels the spin-1/2 case: the critical drive $\eta_c$ decreases monotonically as $\varphi\to\pm\pi$ [see Fig.~\ref{fig:fig3}(a)]. The full mean-field phase diagram at $\varphi=0$ in $(c_+,c_-, \eta)$ space is shown in Fig.~\ref{fig:fig3}(b). Purely bright states appear at $(c_+,c_-,0)=(\pm1,0,0)$ and $(0,0,0)$, while the many-body dark state $\ket{-}^{\otimes N}$ lies at $(0,\pm1,0)$. As $\eta$ increases, oscillatory regions expand from the bright subspace, forming contours with $\tilde\omega_{N_B}\neq 0$, as shown in Figs.~\ref{fig:fig3}(b-c). Introducing $\varphi\neq 0$ rotates the bright and dark subspaces in parameter space without changing their measure, as shown in Figs.~\ref{fig:fig3}(d-e), redistributing the initial state weight between symmetry sectors at fixed $\eta$. This rotation induces transitions between stationary and time-crystalline phases and explains the phase-dependent reduction of the threshold in this three-level model. For sufficiently strong drive ($\eta\gtrsim g/\sqrt{2}$), oscillations occur for all initial states. Although we focus on pure initial states, an extension to mixed states follows by convexity.

\rev{
\subsection{Decoherence-free subspaces in the three-level system} We now examine the effect of finite detuning for the three-level system. Unlike the $\Delta=0$ case, finite detuning introduces qualitatively distinct non-stationary phenomena that are independent of the driving strength $\eta$.

\begin{figure}
    \centering
    \includegraphics[width=0.48\textwidth]{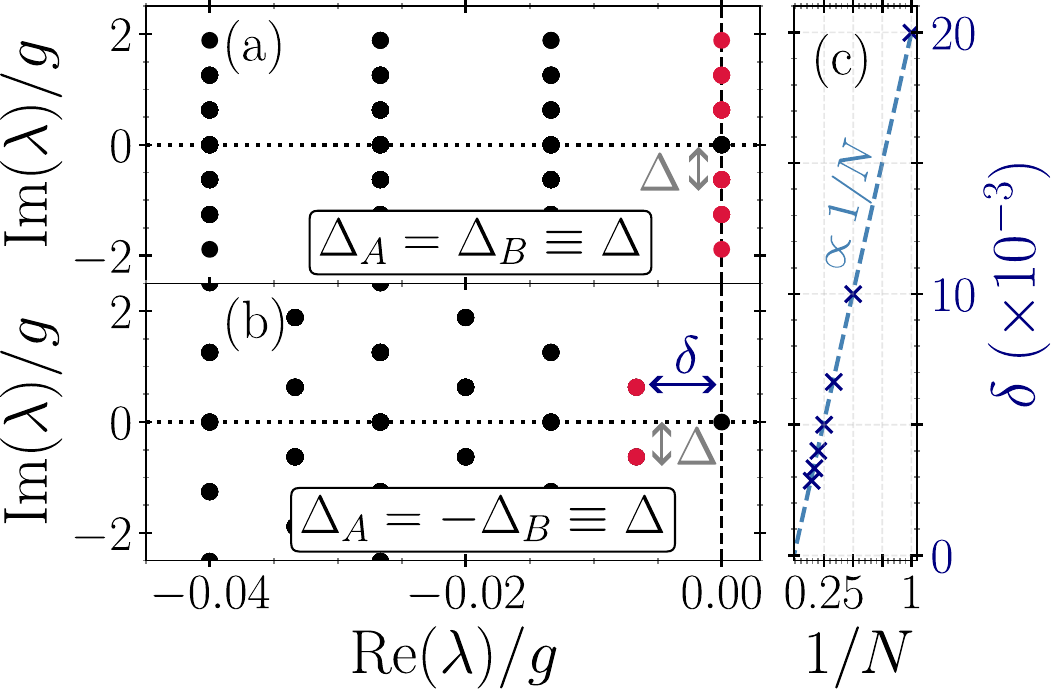}
    \caption{\textbf{Finite and emergent DFS in the three-level system.} Liouvillian spectra for (a) equal detunings $\Delta_A = \Delta_B \equiv \Delta$ and (b) opposing detunings $\Delta_A = -\Delta_B \equiv \Delta$, where red dots denote the smallest real gap. (c) Scaling of the Liouvillian gap $\delta$ versus system size $1/N$. The real gap $\delta$ decreases with system size $N\in\{1,2,\dots,7\}$ (crosses) and can be seen to vanish as $N\to\infty$ (dashed line). The imaginary spectral gap is set by the atomic detuning $\Delta$, which remains constant. Parameters: $\eta/\kappa = 0.025$, $\Delta/\kappa = 0.2\pi$, $g/\kappa = 0.1$}
    \label{fig:fig_dfs_3l}
\end{figure}

 When equal detunings are introduced ($\Delta_A=\Delta_B\equiv\Delta$) the energy splitting between the ground and excited states drives unitary evolution within the bright manifold, generating a DFS marked by Liouvillian eigenmodes with purely imaginary eigenvalues $\lambda_\mathcal{L}=\pm i\Delta$. These correspond to oscillations in the single-particle picture between the $\ket{0}_j$ and $\ket{+}_j$ eigenstates at frequency $\Delta$, and they are clearly visible at finite system sizes, as shown in Fig.~\ref{fig:fig_dfs_3l}(a).

With equal and opposite detunings ($\Delta_A=-\Delta_B\equiv\Delta$), the strong symmetry generator identified above no longer applies, as the detuning pattern creates a new symmetry structure. The effective Hamiltonian now contains terms $\hbar\Delta(\hat{\Lambda}^\dagger_A\hat\Lambda_A - \hat{\Lambda}^\dagger_B\hat\Lambda_B)$ that distinguish between the $A$ and $B$ excited states. At the single-particle level, this leads to the formation of dressed states $\ket{\pm}_j$ with energy splitting $2\Delta$. The jump operator $\hat{\Lambda}_\varphi$ couples predominantly to one combination of these dressed states, while the orthogonal combination becomes increasingly protected as system size grows.

The nonzero real part of the near-zero Liouvillian eigenvalues at finite $N$ arises from weak residual coupling between the protected manifold and the dissipative sector, causing slow decay at rate $\sim \mathcal{O}(1/N)$. As $N \to \infty$, this coupling vanishes and the real part of the gap closes, revealing a true DFS in the thermodynamic limit---an \emph{emergent} DFS (eDFS). Meanwhile, the imaginary gap of size $\Delta$ persists because it stems from the inherent energy splitting between the dressed states, which remains finite regardless of system size. Both features are clearly visible from the Liouvillian spectrum at increasing system sizes, as shown in Fig.~\ref{fig:fig_dfs_3l}(b)--(c).
}

\section{Discussion and Outlook}
In summary, we have shown that a tunable phase provides a practical and efficient control parameter for dissipative phase transitions and long-time dynamics in open quantum systems. Its effect can be understood in terms of a phase-dependent strong symmetry that effectively rotates the invariant symmetry sectors of the Liouvillian, in turn controlling the overlap with symmetry subspaces and substantially reducing the driving threshold for non-stationary dynamics.

We demonstrated the generality of this mechanism in two cavity QED realizations of the Tavis-Cummings model, namely a two-spin-species system and a single-species three-level atomic gas. 
\rev{
Beyond enabling controlled access to emergent dynamical phases, an important consequence of phase-controlled symmetries is that tuning $\varphi$ redistributes the initial-state population across  dynamically decoupled sectors, providing a practical route to symmetry-sector engineering without physical re-preparation of the atomic sample. This identifies laser phase imprinting as a versatile tool for controlling strong-symmetry decompositions in driven-dissipative many-body systems, with direct implications for the preparation and manipulation of symmetry-protected states across a broad class of cavity QED and reservoir-engineered platforms. 

Overall, we envision phase-engineered strong symmetries as a broadly applicable design principle for open quantum systems, where selective switching between protected dark and driven bright states via a single experimental knob opens promising avenues for quantum memories, precision timekeeping, and sensing}.%
\bigskip
\section*{Acknowledgements} The research leading to these results has received funding from the Deutsche Forschungsgemeinschaft (DFG, German Research Foundation) under the Research Unit FOR 5413/1, Grant No. 465199066. P.S. acknowledges support from the Alexander von Humboldt Foundation through a Humboldt research fellowship for postdoctoral researchers. We acknowledge support from the Open Access Publication Fund of the
University of Tübingen.

\section*{Data availability} The data that support the findings of this article are openly available \cite{Zenodo}, with an associated Github repository \cite{Github}.


\appendix

\counterwithout{equation}{section}
\setcounter{equation}{0}
\setcounter{figure}{0}
\setcounter{table}{0}
\renewcommand{\theequation}{A\arabic{equation}}
\renewcommand{\thefigure}{A\arabic{figure}}
\renewcommand{\thetable}{A\arabic{table}}
\renewcommand{\theHequation}{A\arabic{equation}}
\renewcommand{\theHfigure}{A\arabic{figure}}
\renewcommand{\theHtable}{A\arabic{table}}

\bigskip

\section{Microscopic derivation of the models}
\label{app:microscopic}

In this Appendix, we outline all the necessary steps to derive the Hamiltonians introduced in the main text. We start with multilevel atoms that have either two independent ground states or a common ground state, generating the two-species system of two-level atoms or the three-level system, respectively.
\onecolumngrid
\subsection*{Two-species spin-1/2 system} 
First, we consider the two-species two-level atom  system.  The single-particle Hamiltonian for each atomic species coupled to a shared cavity mode with annihilation operator $\hat{a}$ can be written as
\begin{align}
\hat{H} &= \hat{H}_0 + \hat{H}_{at} + \hat{H}_{ac} + \hat{H}_d .
\end{align}
The free Hamiltonian is
\begin{align}
\hat{H}_0 &= \hbar\omega_c\hat a^\dagger \hat a
    +  \hbar\omega_r \ketbra{r}{r}_A + \hbar\omega_s \ketbra{s}{s}_B + \sum_{m=A,B}\hbar\omega_m \ketbra{\uparrow}{\uparrow}_m  ,
\end{align}
where $\omega_c$ is the cavity angular frequency, $\hbar\omega_r,\hbar\omega_s$ are the excited-state energies (for the two Raman intermediate states), and $\hbar\omega_m$ denotes the bare energy of the $\ket{\uparrow}_m$ level for species $m$. Without loss of generality, we have put the energies of the $\ket{\downarrow}_m$ levels to zero. The Raman lasers couple off-resonantly the $\ket{\uparrow}_A$ and $\ket{\uparrow}_B$ states to $\ket{r}_A$ and $\ket{s}_B$, respectively. Their frequencies are given by $\omega_{lr}$ and $\omega_{ls}$, respectively, and the corresponding Hamiltonian describing their action reads,
\begin{align}
\hat{H}_{at} &= \frac{\hbar\Omega_r}{2}e^{-i\omega_{lr}t}\ketbra{r}{\uparrow}_A
         + \frac{\hbar\Omega_s}{2}e^{-i\omega_{ls}t}\ketbra{s}{\uparrow}_B + \text{h.c.},
\end{align}
where $\Omega_{r/s}$ are the corresponding laser Rabi frequencies, which are in general complex-valued (we will keep magnitudes and phases implicit).
The atom--cavity dipole coupling is described by the Hamiltonian
\begin{align}
\hat{H}_{ac} &= \hbar\left( \lambda_r \ketbra{r}{\downarrow}_A\hat a + \lambda_s \ketbra{s}{\downarrow}_B\hat a + \text{h.c.}\right),
\end{align}
with $\lambda_{r/s}$ being the respective cavity coupling constants.
Finally, the cavity is externally driven with drive strength $\eta$ and drive frequency $\omega_{ld}$, and the Hamiltonian describing this reads
\begin{align}
\hat{H}_d &= i\hbar\eta\left( e^{-i\omega_{ld}t}\hat a^\dagger - e^{i\omega_{ld}t}\hat a \right).
\end{align}

We remove explicit time dependence of the Hamiltonian by applying the unitary transformation
\begin{align}
\hat{U}(t) &= e^{i \hat{K} t/\hbar},
\end{align}
where $\hat{K}$ is the diagonal energy operator
\begin{equation}
\hat{K} = \hbar\omega_{ld}\hat a^\dagger\hat a
    + \hbar\omega_{ld}\ketbra{r}{r}_A
    + \hbar\omega_{ld}\ketbra{s}{s}_B  + \hbar(\omega_{ld}-\omega_{lr})\ketbra{\uparrow}{\uparrow}_A
    + \hbar(\omega_{ld}-\omega_{ls})\ketbra{\uparrow}{\uparrow}_B.
\end{equation}
The resulting time-independent Hamiltonian becomes
\begin{equation}\begin{split}
\hat{H}' &= \hbar\Delta_c\hat a^\dagger\hat a
     + \hbar\delta_r\ketbra{r}{r}_A + \hbar\delta_s\ketbra{s}{s}_B
     + \hbar\delta_A\ketbra{\uparrow}{\uparrow}_A + \hbar\delta_B\ketbra{\uparrow}{\uparrow}_B \\
   &\quad +\hbar\left( \frac{\Omega_r}{2}\ketbra{r}{\uparrow}_A + \frac{\Omega_s}{2}\ketbra{s}{\uparrow}_B
          + \lambda_r\ketbra{r}{\downarrow}_A\hat{a} + \lambda_s\ketbra{s}{\downarrow}_B\hat{a} + \text{h.c.}\right)
   + i\hbar\eta(\hat a^\dagger-\hat a) ,
\label{eq:Hprime}\end{split}
\end{equation}
where we have introduced the detunings
\begin{equation}\begin{split}
\Delta_c &=  \omega_c - \omega_{ld}, \quad
\delta_r = \omega_r-\omega_{ld}, \quad
\delta_s = \omega_s-\omega_{ld},\\ 
\delta_A &= (\omega_A-\omega_{ld}) + \omega_{lr}, \quad
\delta_B = (\omega_B-\omega_{ld}) + \omega_{ls}.
\end{split}\end{equation}

We now eliminate $\ket{r}$ and $\ket{s}$ perturbatively under the large single-photon detuning assumptions
\begin{align}
|\delta_r| \gg |\Omega_r|,\; |\lambda_r| \qquad
|\delta_s| \gg |\Omega_s|,\; |\lambda_s|.
\end{align}
Decomposing $\hat{H}'$ into the slow-evolving subspace $\mathcal{S}$, spanned by the cavity field and the low-lying electronic states ($\ket\uparrow_m, \ket\downarrow_m$), and the fast-evolving subspace $\mathcal{F}$, spanned by the far-detuned intermediate states ($\ket{r}, \ket{s}$), the effective Hamiltonian to leading order in $(\delta_{r,s})^{-1}$ is given by the Feshbach-Schur map $$\hat{H}_\mathrm{eff} = \hat{P}_\mathcal{S}\hat{H}'\hat{P}_\mathcal{S}
    - \hat{P}_\mathcal{S}\hat{H}'\hat{P}_\mathcal{F}
      \bigl(\hat{P}_\mathcal{F}\hat{H}'\hat{P}_\mathcal{F}\bigr)^{-1}
      \hat{P}_\mathcal{F}\hat{H}'\hat{P}_\mathcal{S},$$ with $\hat{P}_\mathcal{S}$ and $\hat{P}_\mathcal{F}$ being the projection operators on each subspace. The resulting Hamiltonian (dropping constant energy terms) reads
\begin{equation}
\hat{H}_{\mathrm{eff}} = \hbar\Delta_c\hat a^\dagger\hat a
  + \sum_{m}\left[ (\hbar \delta_m +\hbar V_m)\ketbra{\uparrow}{\uparrow}_m
      + \hbar U_ma^\dagger\hat a\ketbra{\downarrow}{\downarrow}_m +\hbar\left(g_m\hat a\ketbra{\uparrow}{\downarrow}_m + \text{h.c.}\right)\right] +i\hbar\eta\left(\hat{a}^\dagger - \hat{a}\right).
\end{equation}
Here, the single atom-cavity couplings are $g_A \simeq -\dfrac{\lambda_r\Omega_r^*}{2\delta_r}$, $g_B \simeq -\dfrac{\lambda_s\Omega_s^*}{2\delta_s}$, and the leading AC Stark and dispersive shifts are given by
\begin{equation*}
\begin{split}
V_A &\simeq  -\dfrac{|\Omega_r|^2}{4\delta_r}, \quad U_A \simeq -\dfrac{|\lambda_r|^2}{\delta_r}, \\ V_B &\simeq  -\dfrac{|\Omega_s|^2}{4\delta_s}, \quad U_B \simeq -\dfrac{|\lambda_s|^2}{\delta_s}.
\end{split}
\end{equation*}
To obtain the Hamiltonian used in the main text we discard the $U_m$ contributions, assuming operation in the bad-cavity regime where dispersive effects are negligible \cite{Nairn:2025}. Moreover, we choose a relative phase between the Raman lasers,
\begin{equation}
\Omega_r = |\Omega_r|, \quad\Omega_s = |\Omega_r|e^{i\varphi} \label{eq:Raman_phase}
\end{equation}
such that $g_A = |g_A| \equiv g$, $g_B = ge^{-i\varphi}$. Extending this approach to an atomic ensemble via the rescaling $\hat{a}\to\hat{a}/\sqrt{N}$ (which scales the per-atom coupling to $g/\sqrt{N}$ and enhances the drive by $\sqrt{N}$), we arrive at
\begin{equation}
    \hat{H}_{2} = \hbar\Delta_c\hat a^\dagger\hat a + i\hbar\sqrt{N}\eta\left(\hat{a}^\dagger - \hat{a}\right)  + \hbar\sum_m\Delta_m\hat S_m^z+\dfrac{\hbar g}{\sqrt{N}}\left(\hat{a}^\dagger\hat{S}_\varphi + \hat{a}\hat{S}^\dagger_\varphi\right),
\end{equation}
with $\Delta_m \approx \delta_m+V_m$. Here we have used the collective spin operators
$$\hat{S}_{m}^z =  \dfrac{1}{2}\sum_{j=1}^{N/2}\left(\hat{\sigma}^\dagger_{j,m}\hat{\sigma}_{j,m} - \hat{\sigma}_{j,m}\hat{\sigma}^\dagger_{j,m}\right),\; \text{ and }\; \hat{S}_\varphi =\sum_{j=1}^{N/2}\hat{\sigma}_{jA} + e^{i\varphi}\sum_{j=1}^{N/2}\hat{\sigma}_{jB}$$
with $\hat\sigma_{jm}=\ketbra{\downarrow}{\uparrow}_{jm}$. Letting both species have equal detuning $\Delta_A=\Delta_B\equiv\Delta$ we arrive at the form shown in Eq.~\ref{eq:H2}.

\subsection*{Multilevel system with common ground state}
We now focus on deriving the effective Hamiltonian for the collective three-level case. We consider a 5-level system as shown in Fig.~\ref{fig:fig1}(d), which can be seen as a combination of both two-level atomic species sharing a common ground state $\ket{0}$, which maps to an effective spin-1 system. In analogy with the two spin-1/2 systems, the Hamiltonian reads
\begin{equation}
    \hat{H} = \hat{H}_0 + \hat{H}_{at} + \hat{H}_{ac} + \hat{H}_d
\end{equation}
with the free Hamiltonian
\begin{equation}
    \hat{H}_0=\hbar\omega_c\hat{a}^\dagger\hat{a} + \hbar\left(\omega_r\ketbra{r}{r} + \omega_s\ketbra{s}{s}+\omega_B\ketbra{B}{B}+\omega_A\ketbra{A}{A}\right),
\end{equation} 
internal laser coupling of the excited states $\ket{A}$, $\ket{B}$ off-resonantly to the intermediate states $\ket{r}$, $\ket{s}$,
\begin{equation}
  \hat{H}_{at}= \dfrac{\hbar\Omega_r}{2}e^{-i\omega_{lr}t}\ketbra{r}{A}   + \dfrac{\hbar\Omega_s}{2}e^{-i\omega_{ls}t}\ketbra{s}{B}    +\text{h.c.}\quad \text{where } \Omega_r,\Omega_s\in\mathbb{C},
\end{equation}
atom-cavity transitions coupling the intermediate states back to the common ground state $\ket{0}$,
\begin{equation}
    \hat{H}_{ac}= \hbar\left(\lambda_r\ketbra{r}{0}\hat{a} 
    + \lambda_s\ketbra{s}{0}\hat{a}+\text{h.c.}\right),
\end{equation}
and the external cavity driving term
\begin{equation}
    \hat{H}_d = i\hbar\eta\left(e^{-i\omega_{ld}t}\hat{a}^\dagger 
    - e^{i\omega_{ld}t}\hat{a}\right).
\end{equation}

We remove the explicit time dependence by moving to the rotating frame via $\hat{U}(t) = e^{i\hat{K}t/\hbar}$ with
\begin{equation}
\hat{K} = \hbar\omega_{ld}\hat{a}^\dagger\hat{a}
    + \hbar\omega_{ld}\ketbra{r}{r}
    + \hbar\omega_{ld}\ketbra{s}{s}
    + \hbar(\omega_{ld}-\omega_{lr})\ketbra{A}{A}
    + \hbar(\omega_{ld}-\omega_{ls})\ketbra{B}{B},
\end{equation}
yielding the time-independent Hamiltonian 
\begin{equation}
\begin{split}
    \hat{H}' &= \hbar\Delta_c\hat{a}^\dagger\hat{a} 
    + \hbar\delta_r\ketbra{r}{r} + \hbar\delta_s\ketbra{s}{s} 
    + \hbar\delta_A\ketbra{A}{A} + \hbar\delta_B\ketbra{B}{B}\\
    &\quad+ \hbar\left(\frac{\Omega_r}{2}\ketbra{r}{A} 
    + \frac{\Omega_s}{2}\ketbra{s}{B} 
    + \lambda_r\ketbra{r}{0}\hat{a} 
    + \lambda_s\ketbra{s}{0}\hat{a} 
    + \text{h.c.}\right) + i\hbar\eta\left(\hat{a}^\dagger - \hat{a}\right),
\end{split}
\end{equation}
with detunings
\begin{equation}
\begin{split}
\Delta_c &=  \omega_c - \omega_{ld}, \quad
\delta_r = \omega_r-\omega_{ld}, \quad
\delta_s = \omega_s-\omega_{ld},\\ 
\delta_A &= (\omega_A-\omega_{ld}) + \omega_{lr}, \quad
\delta_B = (\omega_B-\omega_{ld}) + \omega_{ls}.
\end{split}
\end{equation}

Following the same adiabatic elimination as in the two-species case under $|\delta_{r,s}|\gg|\Omega_{r,s}|, |\lambda_{r,s}|$, we arrive at the effective Hamiltonian for the 
low-lying manifold $\{|0\rangle, |A\rangle, |B\rangle\}$,
\begin{equation}
\hat{H}_{\rm eff} = \hbar\Delta_c \hat{a}^\dagger \hat{a} 
+ i\hbar\eta(\hat{a}^\dagger-\hat{a}) + \hbar U_0\hat{a}^\dagger \hat{a}\ketbra{0}{0}
+ \sum_{m=A,B} \left[\hbar\Delta_m\ketbra{m}{m}
+ \hbar\left(g_m\hat{a}\ketbra{m}{0} 
+ g_m^*\hat{a}^\dagger\ketbra{0}{m}\right)\right],
\end{equation}
where the effective cavity-mediated coupling constants are
\begin{equation}
g_A \simeq -\frac{\Omega_r^*\lambda_r}{2\delta_r},\qquad 
g_B \simeq -\frac{\Omega_s^*\lambda_s}{2\delta_s},
\end{equation}
the dispersive shift now acts on the common ground state,
\begin{equation}
U_0 \simeq -\frac{|\lambda_r|^2}{\delta_r} - \frac{|\lambda_s|^2}{\delta_s},
\end{equation}
and the dressed detunings are
\begin{equation}
\Delta_A \simeq \delta_A - \frac{|\Omega_r|^2}{4\delta_r}, \qquad
\Delta_B \simeq \delta_B - \frac{|\Omega_s|^2}{4\delta_s}.
\end{equation}
Discarding $U_0$, choosing $\Omega_r = |\Omega_r|$, $\Omega_s = |\Omega_s|e^{i\varphi}$ 
so that $g_A = |g_A|\equiv g$ and $g_B = ge^{-i\varphi}$, and 
extending to $N$ particles via $\hat{a}\to\hat{a}/\sqrt{N}$, we 
recover the Hamiltonian reported in Eq.~(\ref{eq:H3T}),
\begin{equation}
\hat{H}_{3T} =\hbar \Delta_c \hat{a}^\dagger \hat{a} 
+ i\hbar\sqrt{N}\eta(\hat{a}^\dagger-\hat{a}) 
+ \hbar\sum_{m=A,B}\Delta_m\hat{N}_m 
+ \dfrac{\hbar g}{\sqrt{N}}\left(\hat{a}^\dagger\hat{\Lambda}_\varphi 
+ \hat{a}\hat{\Lambda}^\dagger_\varphi\right).
\label{eq:3LS_Ham}
\end{equation}
Here,  we have introduced the collective operators
\begin{equation}
\hat{N}_A=\sum_{j=1}^N\frac{\mathbb{I}_j}{3}+\hat{\lambda}_j^3
+\frac{1}{\sqrt{3}}\hat{\lambda}_j^8, \quad
\hat{N}_B=\sum_{j=1}^N\frac{\mathbb{I}_j}{3}
-\frac{2}{\sqrt{3}}\hat{\lambda}_j^8, \quad 
\hat{\Lambda}_A = \sum_{j=1}^N(\hat{\lambda}^j_1+i\hat{\lambda}^j_2),\quad 
\hat{\Lambda}_B=\sum_{j=1}^N(\hat{\lambda}^j_6+i\hat{\lambda}^j_7)
\end{equation}
with the relation $\hat{\Lambda}_m^\dagger\hat{\Lambda}_m = \hat{N}_m$ labelling the populations in the $m$-th excited leg and the operators $\hat{\Lambda}_m$ responsible for the transition matrix elements $\ketbra{0}{m}$, expressed in terms of the Gell-Mann matrices $\hat{\lambda}_m$ as defined in Eq.~(\ref{eq:gell-mann}) with the phase-dependent collective lowering operator $\hat{\Lambda}_\varphi = \hat{\Lambda}_A + e^{i\varphi}\hat{\Lambda}_B$ (as introduced in the main text).
\bigskip
\twocolumngrid

\section{Effective atom-only description}
\label{app:adiabatic}

In this appendix, we detail the steps required to obtain the effective master equation for the atomic degrees of freedom alone \cite{Azouit:2016}.

In the following, we consider the cavity mode $\hat{a}$ with detuning $\Delta_c$, which couples to a collective atomic operator $\hat{J}$ of $N$ atoms via a Tavis-Cummings type interaction with coupling constant $g/\sqrt{N}$. The cavity is driven with strength $\sqrt{N}\eta$ and decays at rate $\kappa$. For the spin-1/2 model $\hat{{J}} = \hat{S}_\varphi$ whilst for the spin-1 case it is just replaced by $\hat{{J}} = \hat{\Lambda}_\varphi$.

Note we keep the free atomic evolution $\hat{H}_\mathrm{at}$ out of this description as it already lives in the reduced atom-only subspace. The full Liouvillian reads
\begin{equation}
    \dot{\hat{\rho}} = (\mathcal{L}_0 + \mathcal{L}_1)\hat{\rho},
\end{equation}
where
\begin{equation}
\begin{split}
    \mathcal{L}_0\hat{\rho} = -i[\Delta_c \hat{a}^\dagger \hat{a} + i\sqrt{N}(\eta \hat{a}^\dagger - \eta^* \hat{a}), \hat{\rho}] &+ \kappa \mathcal{D}[\hat{a}]\hat{\rho}, \\
    \mathcal{L}_1\hat{\rho} =  \dfrac{-ig}{\sqrt{N}}[\hat{J} \hat{a}^\dagger + \hat{J}^\dagger \hat{a}, \hat{\rho}]&.
\end{split}
\end{equation}

Here $\mathcal{L}_0$ describes the cavity degrees of freedom (free cavity Hamiltonian, coherent drive and decay) and describes the fast dynamics, while $\mathcal{L}_1$ contains the atom-cavity coupling and acts as a slow perturbation. Adiabatic elimination of the cavity is justified when the cavity relaxes much faster than the atomic degrees of freedom, i.e., when the perturbation is small compared to the cavity Liouvillian:
\begin{equation}
    \|\mathcal{L}_1\| \ll \|\mathcal{L}_0\|.
\end{equation}
A convenient explicit condition is
\begin{equation}
    \kappa,|\Delta_c|\gg \frac{g}{\sqrt{N}}.
\end{equation}

In the near-resonant case $(|\Delta_c|\lesssim\kappa)$, this reduces to $\kappa$ being the dominant rate. Under these conditions, the cavity reaches a quasi-steady state on a timescale $\tau_c\sim 1/\max(\kappa,|\Delta_c|)$ and can be adiabatically eliminated by projecting onto the cavity steady state and performing a perturbative expansion in $\mathcal{L}_1$.

To proceed with the elimination and isolate the slow atomic dynamics, we move to a displaced frame defined by
\begin{equation}
\hat{\rho}' = D^\dagger(\alpha)\hat{\rho} D(\alpha),
\quad\text{with}\quad
D(\alpha) = e^{\alpha \hat{a}^\dagger - \alpha^* \hat{a}}.
\end{equation}
In this frame, $D^\dagger(\alpha) \hat{a} D(\alpha) = \hat{a} + \alpha \equiv \delta \hat{a}$. Choosing \begin{equation}
\alpha = \dfrac{\sqrt{N}\eta}{i\Delta_c + \kappa/2}\end{equation} removes the linear drive term and simplifies the Liouvillian to
\begin{equation}
\begin{split}
\mathcal{L}_0'\hat{\rho} &= -i[\Delta_c\, \delta \hat{a}^\dagger \delta \hat{a}, \hat{\rho}] + \kappa \mathcal{D}[\delta \hat{a}]\hat{\rho}, \\
\mathcal{L}_1'\hat{\rho} &= -i g_N [\hat{J}(\alpha^* + \delta \hat{a}^\dagger) + \hat{J}^\dagger(\alpha + \delta \hat{a}), \hat{\rho}],
\end{split}
\end{equation}
where we have defined $g_N=g/\sqrt{N}$. We separate the interaction part into a \emph{coherent drive} and a \emph{fluctuation} term:
\begin{equation}
\mathcal{L}_1' = \mathcal{L}_1^{(\text{drive})} + \mathcal{L}_1^{(\delta)}
\end{equation}
where 
\begin{equation}
\begin{split}
\mathcal{L}_1^{(\text{drive})}\hat{\rho} &= -i[H_{\text{drive}}, \hat{\rho}]\\
\mathcal{L}_1^{(\delta)}\hat{\rho} &= -i g_N [\hat{J} \delta \hat{a}^\dagger + \hat{J}^\dagger \delta \hat{a}, \hat{\rho}].
\end{split}
\end{equation}
for $\quad H_{\text{drive}} = g_N(\alpha^* \hat{J} + \alpha \hat{J}^\dagger)$. To discern between the relevant subspaces, we define the projector $\mathcal{P}$, where
\[
\mathcal{P}\hat{\rho} = \Tr_a[\hat{\rho}] \otimes \ketbra{0}{0},
\]
and its complement $\mathcal{Q} = \mathbb{I} - \mathcal{P}$. The cavity steady state in the displaced frame is the vacuum $\ket{0}$ of $\delta \hat{a}$.
The projector properties imply:
\begin{equation}
\begin{split}
\mathcal{P}\mathcal{L}_1^{(\text{drive})}\mathcal{P} = \mathcal{L}_1^{(\text{drive})}\mathcal{P}, \\
\mathcal{P}\mathcal{L}_1^{(\delta)}\mathcal{P} = 0.
\end{split}
\end{equation}

Now using the standard Nakajima--Zwanzig expansion up to second order \cite{Breuer:2002}, we obtain the dynamics for the reduced density matrix $\hat{\mu}=\mathcal{P}\hat{\rho}$:
\begin{equation}
\dot{\hat{\mu}} = \mathcal{P}\mathcal{L}_1^{(\text{drive})}\mathcal{P}\hat{\rho}
+ \int_0^\infty \! dt \, \mathcal{P} \mathcal{L}_1^{(\delta)} e^{\mathcal{L}_0't} \mathcal{L}_1^{(\delta)} \mathcal{P}\hat{\rho}.
\end{equation}

We now evaluate the terms appearing in this equation. Noting that
\begin{equation}
\mathcal{L}_1^{(\delta)}\hat{\rho} = -i g_N [\hat{V}, \hat{\rho}],
\quad \text{with } \hat{V} = \hat{J} \delta \hat{a}^\dagger + \hat{J}^\dagger \delta \hat{a},
\end{equation}
we can find
\begin{equation}
\begin{split}
\mathcal{L}_1^{(\delta)} e^{\mathcal{L}_0't} \mathcal{L}_1^{(\delta)} \hat{\rho}
= (-i g_N)^2 [\hat{V}, e^{\mathcal{L}_0't}[\hat{V},\hat{\rho}]] \\
= -g_N^2 [\hat{V}, [\hat{V}(t), \hat{\rho}]],
\end{split}
\end{equation}
where $\hat{V}(t) = e^{\mathcal{L}_0't} \hat{V}$.

Expanding the double commutator:
\begin{equation}
\hat{V} \hat{V}(t)\hat{\rho} - \hat{V}\hat{\rho} \hat{V}(t) - \hat{V}(t)\hat{\rho} \hat{V} + \hat{\rho} \hat{V}(t)\hat{V}.
\end{equation}

Expanding $\hat{V}$ and $\hat{V}(t)$ gives:
\begin{equation}
\begin{split}
\hat{V} \hat{V}(t)\hat{\rho} &= (\hat{J} \delta \hat{a}^\dagger + \hat{J}^\dagger \delta \hat{a})(\hat{J} \delta \hat{a}^\dagger(t) + \hat{J}^\dagger \delta \hat{a}(t)) \hat{\rho}, \\
\hat{V}\hat{\rho} \hat{V}(t) &= (\hat{J} \delta \hat{a}^\dagger + \hat{J}^\dagger \delta \hat{a})\hat{\rho} (\hat{J} \delta \hat{a}^\dagger(t) + \hat{J}^\dagger \delta \hat{a}(t)), \\
\hat{V}(t)\hat{\rho} \hat{V} &= (\hat{J} \delta \hat{a}^\dagger(t) + \hat{J}^\dagger \delta \hat{a}(t))\hat{\rho} (\hat{J} \delta \hat{a}^\dagger + \hat{J}^\dagger \delta \hat{a}), \\
\hat{\rho} \hat{V}(t)\hat{V} &= \hat{\rho} (\hat{J} \delta \hat{a}^\dagger(t) + \hat{J}^\dagger \delta \hat{a}(t))(\hat{J} \delta \hat{a}^\dagger + \hat{J}^\dagger \delta \hat{a}).
\end{split}
\end{equation}

Multiplying out gives sixteen terms; we only keep those that survive the trace over the cavity vacuum.

\begin{equation}
\begin{split}
\langle 0|\delta \hat{a}(t)\delta \hat{a}^\dagger|0\rangle &= e^{-(i\Delta_c+\kappa/2)t},\\
\langle 0|\delta \hat{a}^\dagger(t)\delta \hat{a}|0\rangle &= 0.
\end{split}
\end{equation}

After tracing:
\begin{equation}
\begin{split}
\Tr_a[\hat{V} \hat{V}(t)\hat{\rho}] &\to \hat{J} \hat{J}^\dagger e^{-(i\Delta_c+\kappa/2)t} \hat{\mu}_s,\\
\Tr_a[\hat{V}\hat{\rho} \hat{V}(t)] &\to \hat{J} \hat{\mu}_s \hat{J}^\dagger e^{-(i\Delta_c+\kappa/2)t},\\
\Tr_a[\hat{V}(t)\hat{\rho} \hat{V}] &\to \hat{J}^\dagger \hat{\mu}_s \hat{J} e^{(i\Delta_c-\kappa/2)t},\\
\Tr_a[\hat{\rho} \hat{V}(t)\hat{V}] &\to \hat{\mu}_s \hat{J}^\dagger \hat{J} e^{(i\Delta_c-\kappa/2)t}.
\end{split}
\end{equation}
Plugging into the master equation:
\begin{equation}
\begin{split}
\dot{\hat{\mu}}_s
= -g_N^2 &\int_0^\infty dt\, \Big(
[\hat{J}, \hat{J}^\dagger(t)\hat{\mu}_s] \langle \delta \hat{a}(t)\delta \hat{a}^\dagger \rangle\,+ \\
& + [\hat{J}^\dagger, \hat{J}(t)\hat{\mu}_s] \langle \delta \hat{a}^\dagger(t)\delta \hat{a} \rangle
+ \text{H.c.}
\Big)
\end{split}
\end{equation}
Only the first term survives, 
\begin{equation}
\begin{split}
\dot{\hat{\mu}}_s 
= -g_N^2 &\int_0^\infty dt \,\Big[ e^{-(i\Delta_c+\kappa/2)t}\,\times \\
&\left( [\hat{J}, \hat{J}^\dagger \hat{\mu}_s] + [\hat{J}\hat{\mu}_s, \hat{J}^\dagger] \right)\Big].
\end{split}
\end{equation}
Defining the cavity response function
\begin{equation}
   \overline\chi\equiv \int_0^\infty e^{-(i\Delta_c+\kappa/2)t} dt = \frac{1}{\kappa/2 + i\Delta_c},
\end{equation}
the final expression reduces to:
\begin{equation}
\begin{split}
\dot{\hat{\mu}} &= -i[H_\mathrm{at}+H_{\text{drive}} + H_{\text{LS}}, \hat{\mu}]
+ \dfrac{\Gamma}{N}\, \mathcal{D}[\hat{J}]\hat{\mu},
\end{split}
\end{equation}
where
\begin{equation}
\begin{split}
H_{\text{drive}} &=g\eta\bigg(\overline\chi\hat{J} + \overline\chi^*\hat{J}^\dagger\bigg),\\
H_{\text{LS}} &= -\dfrac{g^2}{N}\Delta_c \left|\overline{\chi}\right|^2 \hat{J}^\dagger \hat{J}, \\
\Gamma &= g^2 \kappa\left|\overline{\chi}\right|^2. \label{eq:effective_Lindblad}
\end{split}
\end{equation}
The first term represents the coherent drive from the cavity field amplitude, the second term is the cavity-induced Lamb shift, and the last term describes the effective decay rate $\Gamma$ of the spin subsystems through the lossy cavity.

The validity of this effective description rests on a clear separation of timescales. In particular, the cavity must relax much faster than the atomic dynamics it induces. Once the effective model is obtained, it further demands that all induced atomic rates remain slow compared with the eliminated cavity mode, namely
\begin{equation}
    \bigg\{|g\eta\,\overline\chi|,\,\left|g^2\Delta_c\overline{\chi}^2\right|/N,\,\Gamma \bigg\}\ll \max(\kappa,|\Delta_c|).
\end{equation}
Under these conditions the cavity adiabatically follows its steady state, and the reduced atomic master equation above provides an accurate description of the dynamics. Substituting $\hat{H}_\mathrm{at}=\hbar\Delta\hat S_\varphi^z$ and $\hat{J} = \hat{S}_\varphi$ recovers Eq.~(\ref{eq:2LS_spin_only}), and similarly $\hat{H}_\mathrm{at}=\sum_k\hbar\Delta\hat{N}_k$ and $\hat{J} = \hat{\Lambda}_\varphi$ for Eq.~(\ref{eq:3LS_spinonly}).

\section{Semiclassical equations of motion for the two-species model}
\label{app:eoms}

As stated in the main text, to map out the phase diagram we solve the semiclassical equations of motion obtained under a mean-field ansatz, which become exact in the thermodynamic limit $N\rightarrow\infty$. Using $g_A=|g_A|=g,\,g_B=ge^{-i\varphi}$ they become,
\begin{equation}
    \begin{split}
    \dot {s}_A^x &=-\Delta_As_A^y + igs_A^z\left(\alpha - \alpha^*\right) \\
    \dot {s}_B^x &= -\Delta_B s_B^y  + igs_B^z\left(e^{-i\varphi}\alpha - e^{i\varphi}\alpha^*\right) \\
    \dot {s}_A^y &= \Delta_As_A^x -gs_A^z\left(\alpha^* + \alpha\right) \\
    \dot {s}_B^y &= \Delta_Bs_B^x -gs_B^z\left(e^{i\varphi}\alpha^* + e^{-i\varphi}\alpha\right)\\
    \dot {s}_A^z &= -igs_A^x\left(\alpha-\alpha^*\right) + gs_A^y\left(\alpha^*+\alpha\right)\\
    \dot {s}_B^z &= -igs_B^x\left(e^{-i\varphi}\alpha-e^{i\varphi}\alpha^*\right) + gs_B^y\left(e^{i\varphi}\alpha^*+e^{-i\varphi}\alpha\right)\\
    \dot{\alpha} &= -\left(i\Delta_c +\frac{\kappa}{2}\right)\alpha \\ &- \frac{g}{2}\left(i(s_A^x+e^{i\varphi} s^x_B)+(s_A^y+e^{i\varphi} s^y_B)\right) + \eta \label{eq:eoms12}
    \end{split}
\end{equation}
with $\alpha=\langle\hat{a}\rangle/\sqrt{N}$ and $s^\mu_m = \langle \hat{S}_{m}^\mu \rangle/(N/2)$ for $m=A,B$ where $\hat{S}^{\mu}=\sum_j^{N/2}\hat{\sigma}_j^{(\mu=x,y,z)}$ are the Pauli operators for the collective spins.

\section{Strong-symmetry eigenstates for the two-species model}
\label{app:eigenstates}
\begin{figure}[t!]
    \centering
    \includegraphics[width=0.48\textwidth]{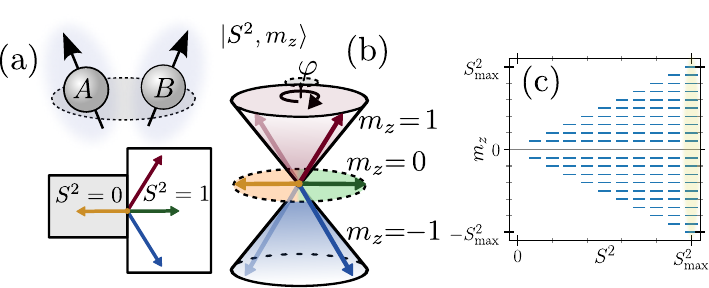}
    \caption{\textbf{Total spin angular momentum eigenstates}. \textbf{(a)} For a generic two spin-1/2 system, the combined state decomposes into a symmetric triplet manifold, $S^2=1$, and an antisymmetric singlet state, $S^2=0$. Within the $m_z=0$ subspace the phase $\varphi$ parametrizes rotations between the symmetric (green) and antisymmetric (orange) eigenstates \textbf{(b)}. When generalizing to higher system sizes $\gg1$, the Hilbert space decomposes into a family of permutation invariant states parametrized by $S^2$ and $m_z$ \textbf{(c)}. For $N_A=N_B$, the rightmost ladder (shaded yellow) corresponds to the fully symmetric (Dicke) subspace, $S^2=N/2$, and the fully antisymmetric subspace corresponds to the unique $S^2=0=m_z$ state.}
    \label{fig:app_angular_mom}
\end{figure}

In this appendix, we expand on the discussion of strong symmetries in the two-species model to further highlight the role of the tunable phase parameter.

With $N_A=N_B=1$, the strong symmetry operator $\mathbf{S}^2_\varphi$ is trivially diagonalized into the eigenbasis
\begin{align}
\ket{T^\varphi} &= \big\{\ket{\downarrow\downarrow},\,\frac{\ket{\uparrow\downarrow}+e^{i\varphi}\ket{\downarrow\uparrow}}{\sqrt{2}},\,\ket{\uparrow\uparrow} \big\}, \nonumber\\
\ket{S^\varphi} &= \frac{\ket{\uparrow\downarrow}-e^{i\varphi}\ket{\downarrow\uparrow}}{\sqrt{2}},
\end{align}
which, for $\varphi=0$, correspond to the usual triplet and singlet angular momentum eigenstates for a system of two qubits \cite{Griffiths:2018}; see Fig.~\ref{fig:app_angular_mom}(a). Varying $\varphi$, there exists within the $m_z=0$ manifold a continuous family of singlet states with orthogonal projection to the symmetric triplet state, see Fig.~\ref{fig:app_angular_mom}(b). The states with $m_z=\pm1$ remain unchanged. Since the dynamics cannot mix distinct sectors, the weights of the initial state in the singlet and triplet manifolds dictate the accessible dynamical pathways and steady states.

The tunable phase $\varphi$ directly shapes this sector decomposition. Consider the initial superposition $\ket{+_x} = \left(\ket{\uparrow\downarrow} + \ket{\downarrow\uparrow}\right)/\sqrt{2}$. Its overlap with the phase-dependent singlet is
\begin{equation}
\langle S^\varphi \mid +_x\rangle = \frac{1-e^{i\varphi}}{2}.
\end{equation}
For $\varphi=0$ this overlap vanishes and $\ket{+_x}$ lies entirely in the symmetric manifold. Tuning $\varphi$ to $\pi$ yields $\ket{S^\pi}=\ket{+_x}$, so the same initial state becomes purely antisymmetric. Thus control of $\varphi$ provides a simple mechanism to steer population between symmetric and antisymmetric sectors.

For a composite system of $N=N_A+N_B\gg1$ spin-1/2 particles, the Hilbert space decomposes into sectors of well-defined total spin $S^2$, ranging from $S^2_{\min}=|N_A - N_B|/2$ to $S^2_{\max}=|N_A + N_B|/2$ \cite{Griffiths:2018}. The maximum sector corresponds to the fully permutationally invariant subspace spanned by Dicke states \cite{Dicke:1954, Carmichael:1980}; see Fig.~\ref{fig:app_angular_mom}(c). The strong symmetry operator $\mathbf{S}^2_\varphi$ continues to distinguish these sectors, and the phase $\varphi$ controls the physical basis states within each by coherently redistributing the overlap of a given initial state among the symmetry sectors. By varying $\varphi$, one can steer the initial state's overlap toward sectors with lower effective spin $S^2$, thereby reducing the system's susceptibility to collective dissipation while keeping the external driving rate unchanged.

\begin{widetext}
\section{Three-level system dynamics: Gell-Mann basis and equations of motion}
\label{app:3level_eoms}

Since the atomic level scheme shares a common level $\vert 0 \rangle$, we describe the system in terms of the Gell-Mann matrices:
\begin{equation}
\begin{split}
\hat{\lambda}_{1} &= \frac{1}{2}
\begin{bmatrix}
0 & 1 & 0 \\ 1 & 0 & 0 \\ 0 & 0 & 0
\end{bmatrix}, \quad
\hat{\lambda}_{2} = \frac{1}{2}
\begin{bmatrix}
0 & -i & 0 \\ i & 0 & 0 \\ 0 & 0 & 0
\end{bmatrix}, \quad
\hat{\lambda}_{3} = \frac{1}{2}
\begin{bmatrix}
1 & 0 & 0 \\ 0 & -1 & 0 \\ 0 & 0 & 0
\end{bmatrix}, \quad
\hat{\lambda}_{4} = \frac{1}{2}
\begin{bmatrix}
0 & 0 & 1 \\ 0 & 0 & 0 \\ 1 & 0 & 0
\end{bmatrix}, \\
\hat{\lambda}_{5} &= \frac{1}{2}
\begin{bmatrix}
0 & 0 & -i \\ 0 & 0 & 0 \\ i & 0 & 0
\end{bmatrix}, \quad
\hat{\lambda}_{6} = \frac{1}{2}
\begin{bmatrix}
0 & 0 & 0 \\ 0 & 0 & 1 \\ 0 & 1 & 0
\end{bmatrix}, \quad
\hat{\lambda}_{7} = \frac{1}{2}
\begin{bmatrix}
0 & 0 & 0 \\ 0 & 0 & -i \\ 0 & i & 0
\end{bmatrix}, \quad
\hat{\lambda}_{8} = \frac{1}{2\sqrt{3}}
\begin{bmatrix}
1 & 0 & 0 \\ 0 & 1 & 0 \\ 0 & 0 & -2
\end{bmatrix}.
\end{split}
\label{eq:gell-mann}
\end{equation}
These matrices are traceless and Hermitian and are the generetors of the SU(3) group. They obey commutation relations $[\hat{\lambda}_a, \hat{\lambda}_b] = i f_{abc} \hat{\lambda}_c$, with the non-zero structure constants $f_{123} = 1$, $f_{147} = f_{246} = f_{257} = f_{345} = 1/2$, $f_{156} = f_{367} = -1/2$, and $f_{458} = f_{678} = \sqrt{3}/2$.

The equations of motion of each Gell-Mann matrix obtained from Eq.~\ref{eq:3LS_Ham} and including cavity dissipation are:

\begin{align}
\dot{\hat{\lambda}}_1 &= -\Delta_A\hat{\lambda}_2
  + ig\Big[ \hat{\lambda}_3(\hat{a} - \hat{a}^\dagger)
  + \frac{e^{i\varphi}\hat{a}^\dagger}{2}(-i\hat{\lambda}_5 + \hat{\lambda}_4)
  + \frac{e^{-i\varphi}\hat{a}}{2}(-i\hat{\lambda}_5 -\hat{\lambda}_4)\Big], \nonumber\\
\dot{\hat{\lambda}}_2 &= \Delta_A\hat{\lambda}_1
  + ig\Big[i\hat{\lambda}_3(\hat{a}^\dagger + \hat{a})
  + \frac{e^{i\varphi}\hat{a}^\dagger}{2}(i\hat{\lambda}_4 +\hat{\lambda}_5)
  + \frac{e^{-i\varphi}\hat{a}}{2}(i\hat{\lambda}_4 -\hat{\lambda}_5)\Big], \nonumber\\
\dot{\hat{\lambda}}_3 &= ig\Big[ -i\hat{\lambda}_2(\hat{a}^\dagger + \hat{a})
  + \hat{\lambda}_1(\hat{a}^\dagger - \hat{a})
  + \frac{e^{i\varphi}\hat{a}^\dagger}{2}(i\hat{\lambda}_7 -\hat{\lambda}_6)
  + \frac{e^{-i\varphi}\hat{a}}{2}(i\hat{\lambda}_7 +\hat{\lambda}_6)\Big], \nonumber\\
\dot{\hat{\lambda}}_4 &= -(\Delta_A-\Delta_B)\hat{\lambda}_5
  + ig\Big[\frac{\hat{a}^\dagger}{2}(i\hat{\lambda}_7 + \hat{\lambda}_6)
  + \frac{\hat{a}}{2}(i\hat{\lambda}_7 - \hat{\lambda}_6)
  + \frac{e^{i\varphi}\hat{a}^\dagger}{2}(-i\hat{\lambda}_2 - \hat{\lambda}_1)
  + \frac{e^{-i\varphi}\hat{a}}{2}(-i\hat{\lambda}_2 + \hat{\lambda}_1)\Big], \nonumber\\
\dot{\hat{\lambda}}_5 &= (\Delta_A-\Delta_B)\hat{\lambda}_4
  + ig\Big[\frac{\hat{a}^\dagger}{2}(-i\hat{\lambda}_6 + \hat{\lambda}_7)
  + \frac{\hat{a}}{2}(-i\hat{\lambda}_6 - \hat{\lambda}_7)
  + \frac{e^{i\varphi}\hat{a}^\dagger}{2}(i\hat{\lambda}_1 - \hat{\lambda}_2)
  + \frac{e^{-i\varphi}\hat{a}}{2}(i\hat{\lambda}_1 + \hat{\lambda}_2)\Big], \nonumber\\
\dot{\hat{\lambda}}_6 &= \Delta_B\hat{\lambda}_7
  + ig\Big[\frac{\hat{a}^\dagger}{2}(i\hat{\lambda}_5-\hat{\lambda}_4)
  + \frac{\hat{a}}{2}(i\hat{\lambda}_5+\hat{\lambda}_4)
  + \frac{1}{2}(e^{-i\varphi}\hat{a}-e^{i\varphi}\hat{a}^\dagger)(\sqrt{3}\hat{\lambda}_8 - \hat{\lambda}_3)\Big], \nonumber\\
\dot{\hat{\lambda}}_7 &= -\Delta_B\hat{\lambda}_6
  + ig\Big[\frac{\hat{a}^\dagger}{2}(-i\hat{\lambda}_4-\hat{\lambda}_5)
  + \frac{\hat{a}}{2}(-i\hat{\lambda}_4+\hat{\lambda}_5)
  + \frac{1}{2}(e^{-i\varphi}\hat{a}+e^{i\varphi}\hat{a}^\dagger)(i\sqrt{3}\hat{\lambda}_8 - i\hat{\lambda}_3)\Big], \nonumber\\
\dot{\hat{\lambda}}_8 &= -i\frac{\sqrt{3}}{2}g\Big[\hat{a}^\dagger e^{i\varphi}(i\hat{\lambda}_7 - \hat{\lambda}_6)
  + \hat{a}e^{-i\varphi}(i\hat{\lambda}_7 + \hat{\lambda}_6)\Big], \nonumber\\
\dot{a} &= \left(i\Delta_c - \frac{\kappa}{2}\right)a
  - ig\Big[(\hat{\lambda}_1 - i\hat{\lambda}_2) + e^{i\varphi}(\hat{\lambda}_6-i\hat{\lambda}_7)\Big] + \eta.
\end{align}
\end{widetext}


The populations in each level are given by 
\begin{equation}
\begin{split}
    \hat{N}_A &= \frac{\hat{I}}{3}+\hat{\lambda}_3+\frac{\hat{\lambda}_8}{\sqrt{3}},\\
    \hat{N}_0 &= \frac{\hat{I}}{3}-\hat{\lambda}_3+\frac{\hat{\lambda}_8}{\sqrt{3}},\\
    \hat{N}_B &= \frac{\hat{I}}{3}-\frac{2\hat{\lambda}_8}{\sqrt{3}}.
    \end{split}
\end{equation}

\subsection*{SU(3) Casimir operators}

For a collective spin-1 system defined by the Gell-Mann matrices, the quadratic Casimir operator is
\begin{equation}
\hat{C}_2 = \sum_{a=1}^8 \hat{\lambda}_a \hat{\lambda}_a,
\end{equation}
with $\langle \hat{C}_2 \rangle = 1/3$ for any pure state in the fundamental representation. Because $\mathfrak{su}(3)$ is a rank-2 algebra, there also exists a cubic Casimir
\begin{equation}
\hat{C}_3 = \sum_{a,b,c} d_{abc} \hat{\lambda}_a \hat{\lambda}_b \hat{\lambda}_c,
\quad d_{abc} = \frac{1}{4}\mathrm{Tr}\left(\{\hat{\lambda}_a,\hat{\lambda}_b\}\hat{\lambda}_c\right),
\end{equation}
with $\langle \hat{C}_3 \rangle = 10/(9\cdot2^3)$. Both Casimirs commute with all generators and are conserved under the dynamics.

\section{Effect of finite cavity detuning}
\label{app:cavity_detuning}

\begin{figure}[t!]
    \centering
    \includegraphics[width=0.48\textwidth]{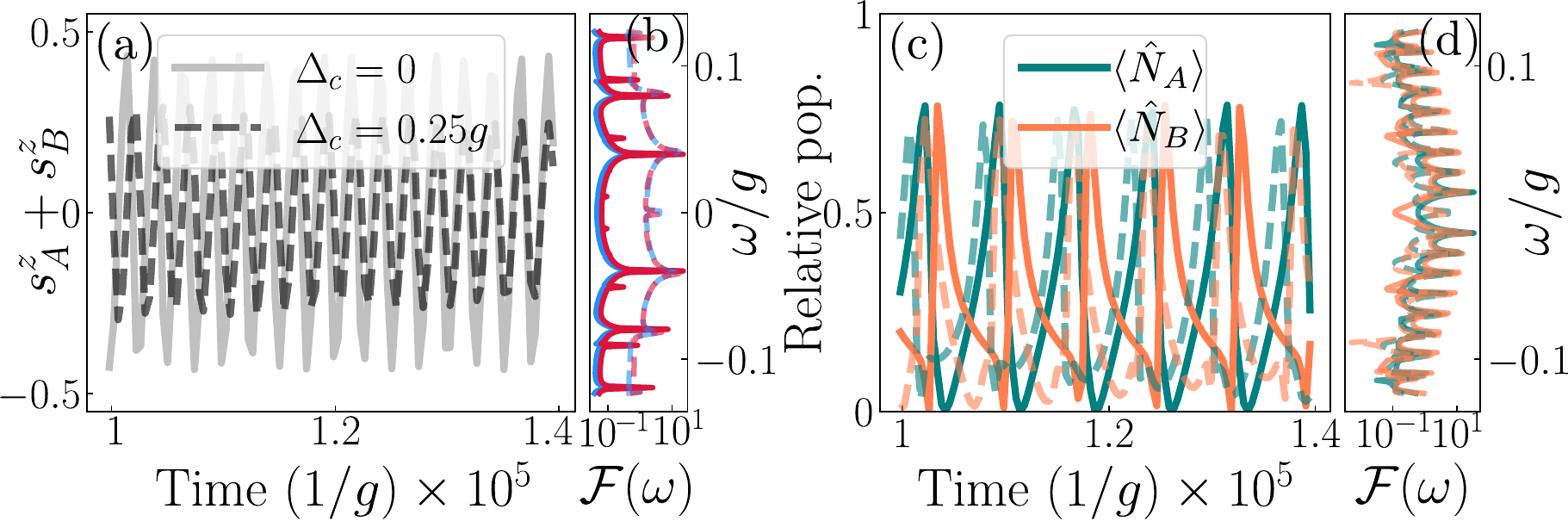}
    \caption{\textbf{Effect of finite cavity detuning.} \textbf{(a)} Semiclassical dynamics of the total magnetization $s_A^z + s_B^z$ deep into the steady state window for the $\Delta_A=\Delta_B=0$, $\varphi = 4\pi/5$, $\eta=0.8g$ point from Fig.~2 in the main text, with and without cavity detuning, and \textbf{(b)} Fourier-domain comparison for each spin species (blue and red) with (dashed) and without (solid) cavity detuning. \textbf{(c)} Dynamics for the excited state populations of the three-level system ($\hat{N}_k=\hat\Lambda_k^\dagger\hat{\Lambda}_k$), for an initial state parametrized with $c_+=0.9/\sqrt{2}$ and $c_-=0.7/\sqrt{2}$, and $\varphi=2\pi/3$. \textbf{(d)} Fourier domain of population dynamics; dashed lines show data with cavity detuning ($\Delta_c/g=0.25$), solid lines for $\Delta_c=0$. The dynamical features and frequency peaks remain largely unchanged with finite $\Delta_c$. In both cases $g/\kappa=0.1$.}
    \label{fig:app_deltac}
\end{figure}

In the main text, we assumed $\Delta_c\ll\kappa$ so that coherent Lamb shifts are suppressed. For completeness, Fig.~\ref{fig:app_deltac} confirms that the inclusion of finite cavity detuning does not alter our findings: the frequency peaks and frequency-matched oscillations remain even at sizeable cavity detunings. Importantly, the sectors that the strong symmetry generator partitions the dynamics into are independent of $\Delta_c$.

\bibliography{resub}

\end{document}